\documentclass[a4paper,11pt]{article}
\usepackage[dvipdfmx]{graphicx}
\usepackage{amsmath}        
\usepackage{amssymb}        
\usepackage{cite}
\usepackage{cancel}
\usepackage{fancybox}
\usepackage{color}
\usepackage{url}
\usepackage{ulem}
\usepackage{pifont}
\usepackage{fancyhdr}
\makeatletter
 
 \@addtoreset{equation}{section}
\makeatother

\def\Sym{{\rm S}}
\def\bfR{{\boldsymbol R}}
\def\bfr{{\boldsymbol r}}

\def\nn{\nonumber\\}
\def\tr{\mathop{\text{tr}}}

\def\slash#1{{\ooalign{\hfil/\hfil\crcr$#1$}}}
\def\half{\frac12}
\def\VEV#1{\left\langle{#1}\right\rangle}

\def\sqr#1#2{{\vcenter{\hrule height.#2pt
      \hbox{\vrule width.#2pt height#1pt \kern#1pt
          \vrule width.#2pt}
      \hrule height.#2pt}}}
\def\bra0{\langle0|}
\def\ket0{|0\rangle}
\def\USp{U\hspace{-0.125em}Sp}
\newdimen\Tdim
\newdimen\Ddim
\def\Tspan#1#2#3{{\setbox0=\hbox{$#3$}% 
\Tdim\ht0\advance\Tdim\dp0\advance\Tdim#1\advance\Tdim#2\Ddim\dp0\advance\Ddim#2\rule[-\Ddim]{0pt}{\Tdim}\box0}}
\def\Onebox#1#2{%
\rule{#1pt}{.#2pt}\hskip-#1pt% lower horizontal line
\rule{.#2pt}{#1pt}\hskip#1pt\hskip-.#2pt% left vertical line
\hskip-#1pt\rule[#1pt]{#1pt}{.#2pt}%  %upper horizontal line
\rule[#1pt]{.#2pt}{.#2pt}\hskip-.#2pt% %upper horizontal line
\rule{.#2pt}{#1pt}\hskip-.#2pt}%                % right vertical line
\def\YGo{\Onebox85\,}
\def\YGoo{\raisebox{-3.5pt}{\Onebox75}\hskip-7pt\raisebox{3.5pt}{\Onebox75}}
\def\YGooo{\raisebox{-7pt}{\Onebox75}\hskip-7pt\raisebox{3.5pt}{\YGoo}}
\def\YGoooo{\raisebox{-7pt}{\YGoo}\hskip-7pt\raisebox{7pt}{\YGoo}\,}
\def\YGt{\Onebox85\Onebox85\,}
\def\YGto{\raisebox{-3.5pt}{\Onebox75}\hskip-7pt%
\raisebox{3.5pt}{\Onebox75\Onebox75}\,}
\def\YGtt{\raisebox{-3.5pt}{\Onebox75\Onebox75}\hskip-14pt%
\raisebox{3.5pt}{\Onebox75\Onebox75}\,}
\def\subYGoo{\raisebox{-1.5pt}{\Onebox34}\hskip-3pt\raisebox{1.5pt}{\Onebox34}}
\def\subYGoooo{\raisebox{-3pt}{\subYGoo}\hskip-3pt\raisebox{3pt}{\subYGoo}}
\def\subYGt{{\Onebox44\Onebox44}}
\def\subYGtt{%
\raisebox{-1.5pt}{\Onebox34\Onebox34}\hskip-6pt
\raisebox{1.5pt}{\Onebox34\Onebox34}}
\def\YI{{\YGoooo}}
\def\subYI{{\subYGoooo}}
\def\YII{{\YGtt}}
\def\subYII{{\subYGtt}}
\def\YIshort{{\YGoo}\,}
\makeatletter
 
 \@addtoreset{equation}{section}
\makeatother
\begin{document}
\thispagestyle{fancy}

\title{Is Symmetry Breaking into Special Subgroup Special?}

\author{Taichiro Kugo\footnote{Electronic address: kugo@yukawa.kyoto-u.ac.jp}
and
Naoki Yamatsu\footnote{Electronic address: yamatsu@gauge.scphys.kyoto-u.ac.jp}
\\
{\it\small
\begin{tabular}{c}
Department of Physics and Maskawa Institute for Science and Culture, \\
Kyoto Sangyo University, Kyoto 603-8555, Japan${}^\ast$\\
Department of Physics, Kyoto University, Kyoto 606-8502,
Japan$^\dag$
\end{tabular}
}
}
\date{\today}

\maketitle

\thispagestyle{fancy}

\begin{abstract}
The purpose of this paper is to show that the symmetry breaking into
 special subgroups is not special at all, contrary to the usual wisdom. 
To demonstrate this explicitly, we examine dynamical symmetry breaking
pattern 
in 4D $SU(N)$ Nambu--Jona-Lasinio type models in which the fermion matter
belongs to an irreducible representation of $SU(N)$. The potential
analysis shows that for almost all cases at the potential minimum the
$SU(N)$ group symmetry is broken to its special subgroups such as
$SO(N)$ or $\USp(N)$ when symmetry breaking occurs. 
\end{abstract}

\section{Introduction}

Symmetries and their
breaking\cite{Nambu:1961tp,Nambu:1961fr,Goldstone:1961eq} are 
important to consider not only the Standard Model (SM) but also
unified theories beyond the SM in particle physics. In the framework of
quantum field theories (QFTs), 
several symmetry breaking mechanisms have been already known, e.g.,
the Higgs mechanism\cite{Higgs:1964pj,Englert:1964et,Guralnik:1964eu}, 
and the dynamical symmetry breaking
mechanism\cite{Nambu:1961tp,Nambu:1961fr,Schwinger:1962tn,Maskawa:1974vs,Maskawa:1975hx,Fukuda:1976zb,Weinberg1976,Susskind:1978ms,Raby:1979my,Dimopoulos:1979es,Farhi:1980xs,Peskin:1980gc,Miransky:1988xi,Miransky:1989ds};
in higher dimensional and string-inspired theories, the Hosotani
mechanism\cite{Hosotani:1983xw,Hosotani:1988bm,Hatanaka:1998yp},
magnetic flux \cite{vonGersdorff:2007uz,Abe:2008fi}
and orbifold breaking mechanism\cite{Dienes:1996yh,Kawamura:1999nj}.

For $SU(n)$ and its breaking via the Higgs mechanism\cite{Li:1973mq},
it is well-known that $SU(n)$ symmetry is broken to $SU(n-1)$ and
$SU(m)\times SU(n-m)\times U(1)$ by the non-vanishing vacuum expectation
value (VEV) of a scalar field in an $SU(n)$ fundamental representation
${\bf n}$ and an $SU(n)$ adjoint representation ${\bf n^2-1}$,
respectively. 
On the other hand, $SU(n)$ symmetry is broken to $SU(n-1)$ or $SO(n)$
and $SU(n-2)$ or $\USp(2\ell) (\ell:=[n/2])$ by the non-vanishing VEV of
a scalar field in an $SU(n)$ 2nd-rank symmetric tensor representation
${\bf n(n+1)/2}$ and an $SU(n)$ 2nd-rank anti-symmetric tensor
representation ${\bf n(n-1)/2}$, respectively.

The above subgroups $SU(n-1)$, $SU(m)\times SU(n-m)\times U(1)$,
$SU(n-1)$, and $SU(n-2)$ are {\it regular subgroups} of
$SU(n)$, while the  $SO(n)$ and $\USp(2\ell)$ are {\it special
subgroups} (or {\it irregular subgroups}) of $SU(n)$
\cite{Dynkin:1957um,Dynkin:1957ek}.
Note that a subgroup $H$ of a group $G$ is called a regular subgroup if
all the Cartan subgroups of $H$ are also the Cartan subgroups of $G$; 
otherwise, the subgroup $H$ is called a special subgroup. 
For example, $SU(2)\times U(1)$ of $SU(3)$ is a regular subgroup, while
$SO(3)\simeq SU(2)$ of $SU(3)$ is a special subgroup. 
If we use the familiar Gell-Mann matrices $\lambda^a$ ($a=1 - 8)$ for
the $SU(3)$ generators, the regular subgroup $SU(2)\times U(1)$ has the
generators $\lambda_1, \lambda_2, \lambda_3, \lambda_8$ when the $SU(2)$
is the usual isospin subgroup, while the generators of the special
subgroup $SO(3)$ are the three anti-symmetric (hence, purely imaginary)
matrices $\lambda_2, \lambda_5, \lambda_7$.
Note that all regular subgroups are obtained by deleting circles
from (extended) Dynkin diagrams, while all special subgroups are not
done so. (For review, see e.g.,
Refs.~\cite{Cahn:1985wk,Slansky:1981yr,Yamatsu:2015gut}.) 

For grand unified theories (GUTs)
in 4 dimensional (4D) theories
\cite{Georgi:1974sy,Fritzsch:1974nn,Gursey:1975ki,Inoue:1977qd,Ida:1980ea,Fujimoto:1981bv,Slansky:1981yr,Georgi:1982jb,Yamatsu:2015gut}
and higher dimensional theories
\cite{Kawamura:2000ev,Kawamura:2000ir,Burdman:2002se,Kim:2002im,Lim:2007jv,Fukuyama:2008pw,Kojima:2011ad,Kawamura:2013rj,Hosotani:2015hoa,Yamatsu:2015rge,Furui:2016owe,Kojima:2016fvv,Kojima:2017qbt,Hosotani:2017ghg,Hosotani:2017edv},
a lot of GUT models use the Lie groups and their {\it
regular subgroups} in a series:
\begin{align}
E_6\supset SO(10)\supset SU(5)\supset G_{\rm SM},
\end{align}
where $G_{\rm SM}:=SU(3)_C\times SU(2)_L\times U(1)_Y$, and
we omitted several $U(1)$ subgroups.
A few GUT models
\cite{Yamatsu:2017sgu,Yamatsu:2017ssg,Yamatsu:2018fsg} 
are known to use not only the {\it
regular subgroups} but also {\it special subgroups}
such as
\begin{align}
SO(32)\supset SU(16)\supset SO(10)\supset SU(5)\supset G_{\rm SM},
\end{align}
where we omitted several $U(1)$ subgroups for regular subgroups.
The $SU(16)$ group has a maximal special subgroup $SO(10)$, {\bf 16} 
spinor of which is 
identified with the defining {\bf 16} representation of $SU(16)$.  
The $SU(16)$ symmetry can be broken to $SO(10)$ via the VEV of
the $SU(16)$ ${\bf 5440}$ representation corresponding to a Young tableau
$\YII$.
Note that a subgroup $H$ of $G$ is called maximal if there is no
larger subgroup containing it except $G$ itself.
For example, $U(1)\times U(1)$ of $SU(3)$ is not a maximal subgroup
because one of $U(1)$ is contained in 
the regular subgroup $SU(2)\subset SU(3)$.
Some typical examples of the maximal special subgroups of
$SU(n)$ are listed in Table~\ref{table:SUn_maximal_subgroup}.

\begin{table}[thb]
\begin{center}
\begin{tabular}{cll}
\hline
Rank& $G\supset H$&Condition\\\hline
$mn-1$ &$SU(mn)\supset SU(m)\times SU(n)$&$(m,n\geq2)$\\
$2n$   &$SU(2n+1)\supset SO(2n+1)$ &$(n\geq1)$\\
$2n-1$ &$SU(2n)\supset\USp(2n), SO(2n)$ &$(n\geq2)$\\
 $\frac{(n+1)(n-2)}{2}$
 &$SU\left({\frac{(n(n-1)}{2}}\right)\supset SU(n)$ &$(n\geq3)$\\
$\frac{(n-1)(n+2)}{2}$&$SU\left({\frac{n(n+1)}{2}}\right)\supset SU(n)$ &$(n\geq2)$\\
$15$  &$SU(16)\supset SO(10)$ \\
$26$  &$SU(27)\supset E_6$ \\\hline
\end{tabular}
\end{center}
\caption{$H$ is a maximal special subgroup of $G=SU(N)$. This table is a part of
Tables 4 and 5 of Ref.~\cite{Yamatsu:2015gut}. 
Note that this table is not a complete list of maximal special subgroups.}
\label{table:SUn_maximal_subgroup}
\end{table}

When we discuss spontaneous symmetry breaking, it is important to know 
not only subgroups but also {\it little groups}. 
A little group $H_\phi$ of a 
vector $\phi$ in a representation $\bfR$ of $G$ is defined by 
\begin{equation}
H_\phi:= \Bigl\{ g \ \Big| \ g\phi=\phi,\ g\in G\ \Bigr\}.
\label{eq:LittleGroup}
\end{equation}
This {\it little group} $H_\phi$ of $G$ depends not only on the
representation $\bfR$ of $\phi$ but also the vector (value) $\phi$
itself. The vector $\phi$ must be an $H_\phi$-singlet, so that a
subgroup $H$ can be a little group of $G$ for some representation $\bfR$
only when $\bfR$ contains at least one  $H$-singlet.
For example, the maximal little groups of $SU(3)$ ${\bf 3}$, ${\bf 6}$,
and ${\bf 8}$ representations are $SU(2)$(R), $SU(2)$(R) and $SO(3)$(S),
and $SU(2)\times U(1)$(R), where (R) and (S) stand for regular and special
subgroups, respectively. Practically, the so-called Michel's conjecture 
\cite{Michel:1980pc} are very useful.
The Michel's conjecture tells us that a potential that consists of a
scalar field in an irreducible representation $\bfR$ of 
a group $G$ has its
potential minimum that preserves one of its maximal little groups $H$ 
of $\bfR$.
This conjecture drastically reduces the number of states especially for
higher rank group cases.

Many people vaguely believe that symmetry groups are broken to
only {\it regular} subgroups, not to {\it special} subgroups.
The main purpose of this paper is to show that
symmetry breaking into {\it special} subgroups are not special
by using 4D Nambu--Jona-Lasinio (NJL) type model in the framework of
dynamical symmetry breaking scenario
\cite{Kugo:1994qr}.

This paper is organized as follows. In Sec.~\ref{Sec:NJL_model}
we first review a 4D NJL type model to show the method of potential
analysis. 
In Secs.~\ref{Sec:NJL_SUn-def} and \ref{Sec:NJL_SUn-Anti},
we apply the method for two cases in which the fermion belongs to the
defining representation and rank-2 anti-symmetric representations of
$SU(n)$, respectively. For the latter NJL model with rank-2 
anti-symmetric fermion, we will show, in particular, that $SU(16)$
symmetry breaks into two degenerate vacua of special subgroups $SO(16)$
and $SO(10)$ for a certain region of coupling constants. However, 
this degeneracy actually turns out to cause the mixing of the two 
vacua and leads to the total breaking of the $SU(16)$ symmetry,
generally. Some detailed identification of the scalar VEVs is necessary
to discuss this mixing phenomenon of the degenerate vacua, so the task
will be given in the Appendix.
Section~\ref{Sec:Summary} is devoted to a summary and discussions, 
where we 
also note the similarity of the present results to the previous one in 
Ref.~\cite{Kugo:1994qr} for the 
$E_6$ NJL model with fundamental {\bf27} fermion.

\section{Nambu--Jona-Lasinio type model}
\label{Sec:NJL_model}

We consider a 4D Nambu-Jona-Lasinio (NJL) type model
\cite{Nambu:1961tp,Nambu:1961fr}
in which the fermion
matter $\psi=(\psi_I)$ ($I=1,2,\cdots,{\rm dim}\bfR=d$) belongs to an
irreducible representation $\bfR$ of dimension $d$ of the group $G$. The
each fermion field $\psi_I$ is the two-component left-handed spinor
$\psi_{I\alpha}$ with an undotted spinor index $\alpha$ running over 1
to 2. Then the Lorentz scalar fermion bilinears
$\psi_I\psi_J :=\psi_I^\alpha\psi_{J\alpha}
:=\varepsilon^{\alpha\beta}\psi_{I\beta}\psi_{J\alpha}$ and 
$\bar\psi^I\bar\psi^J
:=\bar\psi^I_{\dot\alpha}\bar\psi^{J\dot\alpha} = (\psi_I\psi_J)^*$ 
are symmetric under exchange $I \leftrightarrow J$ 
owing to the Fermi statistics of $\psi_I$. 
Assume that the
symmetric tensor product $(\bfR\times\bfR)_\Sym$ is decomposed into
$n_R$ irreducible representations $\bfR_p$: 
\begin{equation}
(\bfR\times\bfR)_\Sym  = \sum_{p=1}^{n_R} \, \bfR_p . 
\end{equation} 
Then the NJL Lagrangian has $n_R$ independent 4-fermion interaction
terms: 
\begin{align}
{\cal L}=& 
\bar\psi^I\,i\overline{\sigma}^\mu\partial_\mu\psi_I
+ \sum_{p=1}^{n_R} \frac14{G_{\bfR_p}} 
 \bigl(\psi_I\psi_J\bigr)_{\bfR_p}
 \bigl(\bar\psi^I\bar\psi^J\bigr)_{\overline\bfR_p},
\end{align}
where $\bigl(\psi_I\psi_J\bigr)_{\bfR_p}$ denotes the projection of the
fermion bilinear into the irreducible component $\bfR_p$. 
Introducing auxiliary complex scalar fields 
$\Phi_{\bfR_p}$ ($p=1,\cdots,n_R$) standing for each of the irreducible
components $-(G_{\bfR_p}/2)\bigl(\psi_I\psi_J\bigr)_{\bfR_p}$
\cite{Gross:1974jv,Kugo:1978ct},
we rewrite this Lagrangian into
\begin{align}
{\cal L}&=
\bar\psi^I i\bar{\slash{\partial}}\psi_I
-\sum_{p=1}^{n_R}\left\{\half 
\bigl(\psi_I\psi_J\bigr)_{\bfR_p}\Phi_{\bfR_p}^{\dag IJ}
+\half\bigl(\bar\psi^I\bar\psi^J\bigr)_{\overline\bfR_p}\Phi_{\bfR_pIJ}
+M^2_{\bfR_p}\,\Phi_{\bfR_p}^{\dag\,IJ}\!\Phi_{\bfR_p\,IJ}\right\}
\nonumber\\
&=
\bar\psi^I i\bar{\slash\partial}\psi_I
-\half\psi_I\psi_J \Phi^{\dag\,IJ}
-\half\bar\psi^I\bar\psi^J \Phi_{IJ}
-\sum_{p=1}^{n_R} 
M^2_{\bfR_p}\,\tr \bigl(\Phi_{\bfR_p}^{\dag}\Phi_{\bfR_p}\bigr),
\end{align}
where $\bar{\slash{\partial}}:=\bar\sigma^\mu\partial_\mu$, 
$M^2_{\bfR_p}:=1/G_{\bfR_p}$, $\Phi_{IJ}$ without
irreducible index $\bfR_p$ was introduced in 
the second line to denote the sum
\begin{align}
\Phi_{IJ} = \sum_{p=1}^{n_R} \Phi_{\bfR_p IJ},
\end{align}
which now stands for the general symmetric $d\times d$ complex matrix
with no more constraint. Now, noting that the kinetic and Yukawa terms
of the fermion can be rewritten into 
%%%%%%%%%%%%%%%%%%%%%%%%%%%
\def\McalH{
\begin{pmatrix}
\Phi^\dagger&   \\
    & \Phi\\
\end{pmatrix}}
\def\Mder{
\begin{pmatrix}
& i\slash{\partial}\  =i\sigma^\mu\partial_\mu\\
i\bar{\slash{\partial}} &  \\
\end{pmatrix}}
%%%%%%%%%%%%%%%%%%%%%%%%%%%%
\begin{align}
&\bar\psi^I i\bar{\slash\partial}\psi_I
-\half\psi_I\psi_J \Phi^{\dag\,IJ}
-\half\bar\psi^I\bar\psi^J \Phi_{IJ} \nn
&=
\frac12
 \begin{pmatrix}\psi_I^{\alpha}&
 \overline{\psi}_{\dot{\alpha}}^I \end{pmatrix}
\begin{pmatrix}
-\Phi^{\dagger IJ}\delta_\alpha^\beta&
i(\slash{\partial})_{\alpha\dot{\beta}}\delta_J^I\\
i(\slash{\overline{\partial}})^{\dot{\alpha}\beta}\delta_I^J&
-\Phi_{IJ}\delta_{\dot{\beta}}^{\dot{\alpha}}\\
\end{pmatrix}
\begin{pmatrix}
\psi_{\beta J}\\
\overline{\psi}^{\dot{\beta}J}
\end{pmatrix}
\nn
&=
\frac{1}{2}
\bar\Psi_I\left( i\gamma^\mu\partial_\mu- {\boldsymbol \varPhi} 
\right)_{IJ}\Psi_J
\nn
&\hbox{with} \quad 
 i\gamma^\mu\partial_\mu:=\Mder, \quad
 {\boldsymbol \varPhi}:=\McalH, \quad 
\Psi_I:= 
\begin{pmatrix}
\psi_I\\
\overline{\psi}^I
\end{pmatrix}
\end{align} 
up to the total derivative terms, one can calculate the effective
potential in the leading order in $1/N$ \footnote{We regard each
$\psi_I$ as $N$-plet of a certain fictitious `color' group $U(N)$ and do
the expansion in $1/N$. We, however, set $N=1$.} 
as
\begin{align}
 V^{\text{leading}}(\Phi)&=
 V^{\text{tree}}(\Phi)+V^{\text{1-loop}}(\Phi),\nn
 V^{\text{tree}}(\Phi) &=
 \sum_{p=1}^{n_R} M^2_{\bfR_p}\,
 \tr \bigl(\Phi_{\bfR_p}^{\dag}\Phi_{\bfR_p}\bigr),\nn
 V^{\text{1-loop}}(\Phi)&=
 -\frac{\hbar}{2}\int^\Lambda{d^4p\over i(2\pi)^4} \ln \det_{4\otimes d}
 \left( \gamma^\mu p_\mu-{\boldsymbol \varPhi} \right)
 \nonumber\\
 &=-\frac{\hbar}{2}\int^\Lambda{d^4p\over i(2\pi)^4}{\ln}\left[ \det_{4\otimes d}
\left( \gamma^\mu p_\mu\right)\cdot \det_{4\otimes d}
\left( 1- (\gamma^\mu p_\mu)^{-1}{\boldsymbol \varPhi} \right)\right]
\nonumber\\
&=-\frac{\hbar}{2}\int^\Lambda{d^4p\over i(2\pi)^4}{\ln}\left[ \det_{4\otimes d}
\left( \gamma^\mu p_\mu\right)\cdot \det_{4\otimes d}{}^{1/2}
\left( 1- (\gamma^\mu p_\mu)^{-1}{\boldsymbol \varPhi} \right)
\left( 1+ (\gamma^\mu p_\mu)^{-1}{\boldsymbol \varPhi} \right)\right],
\end{align}
where $\det_{4\otimes d}$ denotes the determinant of 
$4d\times4d$ matrix. Now inserting
\begin{align}
[(\gamma^\mu p_\mu)^{-1}{\boldsymbol \varPhi}]^2
        &=
\left[\frac{1}{p^2}
\begin{pmatrix}
&\slash{p}\\
\slash{\bar{p}}&\\
\end{pmatrix}
\begin{pmatrix}
 \Phi^\dagger&     \\
  & \Phi 
\end{pmatrix}
\right]^2 
=
\frac{1}{p^2}
\begin{pmatrix}
{\Phi\Phi^\dagger}& \\
& \Phi^\dagger\Phi 
\end{pmatrix},
\end{align}
the 1-loop potential part reads 
\begin{align}
V^{\text{1-loop}}(\Phi)
&=
-\frac{\hbar}{2}\int^\Lambda{d^4p\over i(2\pi)^4}
\frac12 \ln \det_{4\otimes d}
\begin{pmatrix}
p^2-{\Phi}\Phi^\dagger&\\
&p^2-\Phi^\dagger{\Phi}\\
\end{pmatrix}
\nonumber\\
=&
-\hbar\int^\Lambda{d^4p\over i(2\pi)^4}
\ \tr_d \ln
\left(p^2-{\Phi^\dagger}\Phi\right) \quad\ \Big(\because 
\det_d(p^2-\Phi{\Phi^\dagger})=\det_d(p^2-{\Phi^\dagger}\Phi)\Big),
\end{align}
where $\tr_d$ denotes the trace of $d\times d$ matrix and the last
relation follows 
from $\Phi\Phi ^\dagger=(\Phi^\dagger\Phi)^{\rm T}$ since $\Phi$ is a
symmetric matrix. Since the $\Phi$-independent constant
$V^{\text{1-loop}}(0)$ can be discarded for our purpose finding the
potential minimum, we henceforth redefine the 1-loop part
$V^{\text{1-loop}}(\Phi)$ actually to be
$V^{\text{1-loop}}(\Phi)-V^{\text{1-loop}}(0)$ by subtracting it. 
Then, if we define the loop momentum integration by imposing the 
UV cutoff $\Lambda$ on the Euclideanized momentum
$p^0\rightarrow ip^4_{\rm E}$ as $p_{\rm E}^2<\Lambda^2$, we have the
formula 
\begin{align}
&\hspace{-1em}\int^\Lambda{d^4p\over i(2\pi)^4} 
\left(\ln\Bigl(-p^2 +m^2 \Bigr) 
-\ln\Bigl(-p^2\Bigr)\right) := 
\int_{p^2_{\rm E}\leq\Lambda^2}\frac{d^4p^{\phantom 3}_{\rm E}}{(2\pi)^4} 
\ln\Bigl(\frac{p_{\rm E}^2 +m^2}{p_{\rm E}^2} \Bigr) \nn 
&={1\over32\pi^2}\left\{
\Lambda^4\ln\Bigl(1+\frac{m^2}{\Lambda^2}\Bigr) 
-m^4\ln\Bigl(1+\frac{\Lambda^2}{m^2}\Bigr)+m^2\Lambda^2
\right\} =: f(m^2).
\end{align}
This formula is valid even when $m^2$ is a general Hermitian matrix if
$f(m^2)$ is understood to be a matrix-valued function of the matrix.    
So the final form of the 1-loop part is
\begin{align}
V^{\text{1-loop}}
&=-  \tr_d f(\Phi^\dagger\Phi)  
=-  \sum_{I=1}^d f(m^2_I),
\end{align}
where $m^2_I$ are $d$ eigenvalues of the Hermitian matrix
$\Phi^\dagger\Phi$, which stand for $d$ mass-square eigenvalues of the
fermion $\psi_I$. 

\begin{figure}[tbh]
\begin{center}
   \includegraphics[bb=0 0 402 246,height=5cm]{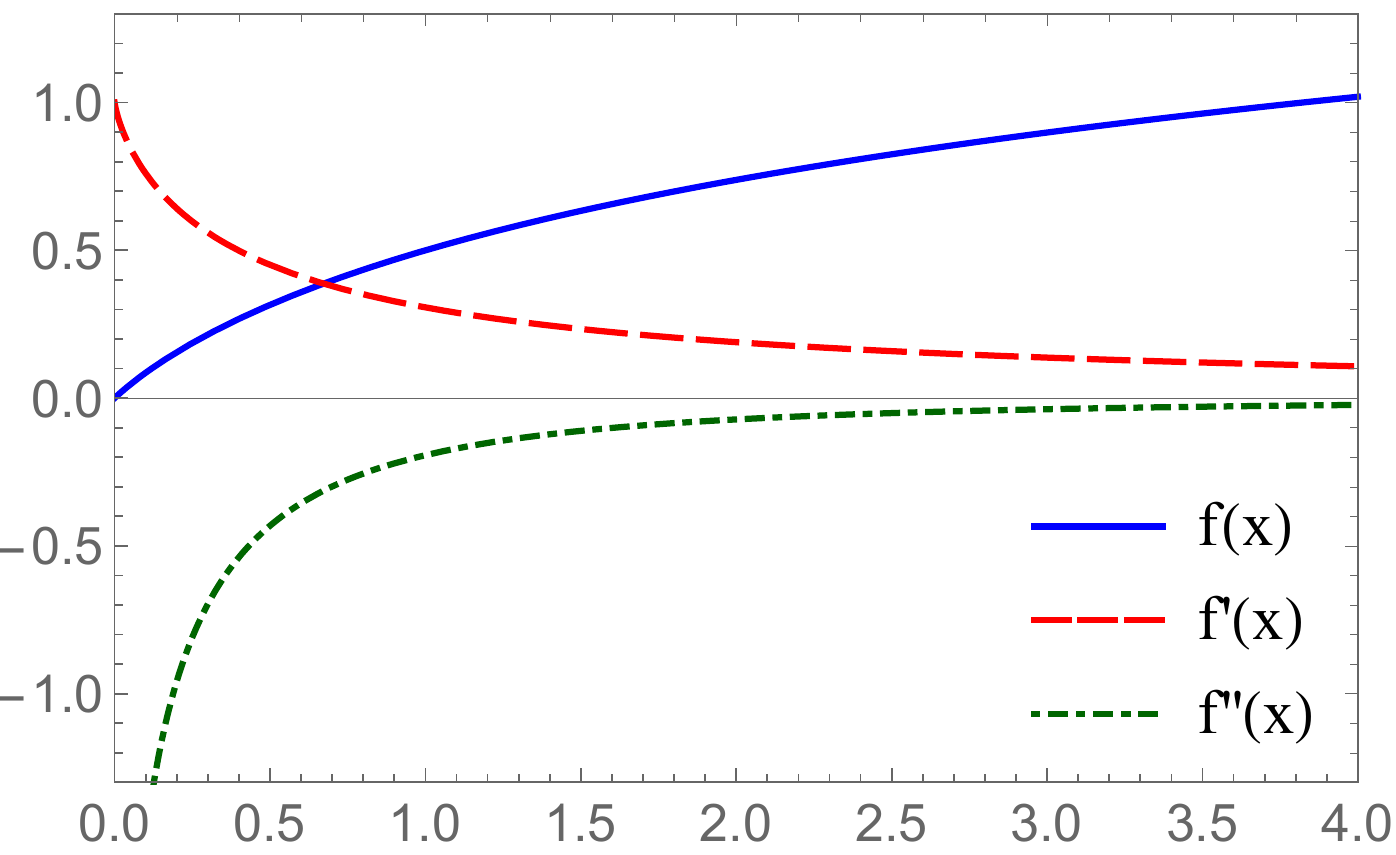}
 \caption{The figure shows the behavior of the functions
 $\bar f(x),  \bar f'(x)$ and  $\bar f''(x)$ defined in
 Eqs.~(\ref{Eq:fx})-(\ref{Eq:fx''}), where $f(x), f'(x), f''(x)$ in the
 figure  stand  for $\bar f(x), \bar f'(x)$, $\bar f''(x)$,
 respectively.} 
 \label{fig:fx-functions}
\end{center}
\end{figure}

This 1-loop function $f(m^2)$ is monotonically increasing 
upward-convex function. 
In Fig.~\ref{fig:fx-functions}, we plot the rescaled dimensionless
function 
$\bar f(x) :=\frac{16\pi^2}{\Lambda^4}f(m^2)$ of $x=m^2/\Lambda^2\geq0$ 
as well as the first and second derivatives:
\begin{align}
\bar f(x) 
&= \frac12 
\left(\ln(1+x)-x^2\ln\Bigl(1+\frac1x\Bigr)+x\right) \simeq  
\left\{
\begin{array}{lcl}
x +\frac{x^2}2\ln x& \text{for} & x\ll1  \\
\frac12\ln x +\frac14 & \text{for} & x\gg1
\end{array}
\right.,
\label{Eq:fx}\\
\bar f'(x) 
&= 1- x\ln\Bigl(1+\frac1x\Bigr)
\simeq  
\left\{
\begin{array}{lcl}
1 + x\ln x& \text{for} & x\ll1  \\
\frac1{2x} & \text{for} & x\gg1
\end{array}
\right.,
\label{Eq:fx'}\\ 
\bar f''(x) 
&= \frac1{1+x}- \ln\Bigl(1+\frac1x\Bigr).
\label{Eq:fx''} 
\end{align}
They lead to $\bar f'(x)>0$, $\bar f''(x)<0$ 
in the whole region $x>0$.
For the single component $\Phi\Phi ^\dagger=v^2$ case the leading
potential is given by 
\begin{equation}
V^{\text{leading}}(v^2)=M^2 v^2- f(v^2) 
= \Lambda^2\Bigl(M^2x-{\Lambda^2\over16\pi^2}\bar f(x)\Bigr).\qquad \bigl(x:=\frac{v^2}{\Lambda^2}\bigr)
\end{equation}
From the behavior of $\bar f(x)$ in Fig.~~\ref{fig:fx-functions}, we see
that the critical coupling constant $G_{\text{crt}}=M^{-2}_{\text{crt}}$
for $d ={\rm dim}\bfR=1$ case is given by
\begin{equation}
{M^{2}_{\text{crt}}\over\Lambda^2} =\frac1{16\pi^2}\quad \rightarrow\quad 
G_{\text{crt}}={16\pi^2\over\Lambda^2}\,,
\end{equation}
as determined by the decreasing
condition of the function $V^{\text{leading}}\propto M^2x - f(x) \simeq 
(M^2-\Lambda^2/16\pi^2)x\ (x\ll\Lambda^2)$ 
around $x=0$.

It is convenient to rewrite the tree part potential into the following
form by picking up one particular representation, say $\bfR_1$, from
$\bfR_p$'s:  
\begin{align}
V^{\text{tree}} &=
M^2_{\bfR_1}\,\tr \bigl(\Phi^\dagger\Phi\bigr) +
\sum_{p=2}^{n_R} (M^2_{\bfR_p}-M^2_{\bfR_1})\,\tr 
\bigl(\Phi_{\bfR_p}^{\dag}\Phi_{\bfR_p}\bigr).
\label{eq:11}
\end{align}
This is because $\Phi=\sum_p \Phi_{\bfR_p}$ is the general
(unconstrained) symmetric 
$d\times d$ matrix which solely appears in the 1-loop part potential
$V^{\text{1-loop}}$, while $\Phi_{\bfR_p}$'s are constrained matrices
subject to non-trivial condition belonging to the irreducible
representation $\bfR_p$, so satisfying  the orthogonality
$\tr (\Phi^\dagger_{\bfR_p}\Phi_{\bfR_{p'}})=0$ for $p\not=p'$.

Whether a symmetry breaking pattern $G\rightarrow H$ is possible or
not is found as follows. Expand each $G$-irreducible representation 
$\bfR_p$ into $H$-irreducible components $\bfr_p^{H(i)}$:
\begin{equation}
\bfR_p = \sum_{i=1}^{n^H_p} \bfr_p^{H(i)}.
\end{equation}  
If there is an $H$-singlet contained in this decomposition for one $p$
or more, then the possibility for the breaking $G\rightarrow H$ exists. 
So assuming the non-zero VEV for all the $H$-singlets and identifying 
how those singlet VEV('s) is contained in the scalars $\Phi_{\bfR_p}$,
we can calculate the potential and find the potential values at the
minimum points of the potential. We do this calculation for all
possibilities of the subgroup $H$.
Then we can find the true minimum, comparing those minimum values for
all possible choices of $H$. 
To find the symmetry breaking that realizes the lowest minimum of the potential, 
we should note that the present potential $V(\Phi)$ in Eq.(2.6) consists 
of negative definite 
monotonically decreasing 1-loop potential 
$V^{\text{1-loop}}(\Phi^\dagger\Phi)=-\sum_I f(m_I^2)$ 
and positive definite tree potential 
$\sum_p M^2_{\bfR_p}\tr(\Phi^\dagger_{\bfR_p}\Phi_{\bfR_p})$. So,  
to realize the lower values of the potential, it is preferable that 
\begin{eqnarray}
&&\hbox{
1. the number of massive fermions $\psi_I$ with $m_I^2\not=0$ is 
as large as possible}.  \nn
&&\hbox{
2. the condensation (nonvanishing VEV) $\VEV{\Phi_{\bfR_p}}$ occurs in 
the direction of} \nn 
&&\hspace{1.5em}\hbox{stronger coupling channel ${\bfR_p}$, i.e., 
with smaller $M^2_{\bfR_p}$}.
\label{eq:criterion}
\end{eqnarray}
To examine all the possibilities systematically, we consider all the 
maximal little groups for every $\Phi_{\bfR_p}$ where the maximal little groups 
of $\Phi_{\bfR_p}$ are defined as follows: The little group of the VEV 
$\VEV{\Phi_{\bfR_p}}$ of the group $G$ is 
$H_{\VEV{\Phi_{\bfR_p}}}$ defined in Eq.~(\ref{eq:LittleGroup}) 
for the vector $\phi=\VEV{\Phi_{\bfR_p}}$, so that
the VEV $\VEV{\Phi_{\bfR_p}}$ belongs to an 
$H_{\VEV{\Phi_{\bfR_p}}}$-singlet. As the VEV $\VEV{\Phi_{\bfR_p}}$
changes, the little group 
$H_{\VEV{\Phi_{\bfR_p}}}$ also changes. A little group $H$ of some VEV 
$\VEV{\Phi_{\bfR_p}}_0$
is called 
{\it maximal little group} of $\Phi_{\bfR_p}$ if there are no VEV $\VEV{\Phi_{\bfR_p}}$ whose 
little group $H_{\VEV{\Phi_{\bfR_p}}}$ satisfies $G\supset H_{\VEV{\Phi_{\bfR_p}}}\supset H$. 
For certain systems of restricted class of potentials of scalar fields, 
there is Michel's conjecture\cite{Michel:1980pc,Slansky:1981yr} which
claims that the group symmetry can breaks down only to one of the
maximal little groups of the considered scalar field
$\Phi_{\bfR_p}$. Our system does 
not fall into such a restricted system, so that the lowest potential
needs not be realized by one of the maximal little groups. But we can
anyway consider the breaking possibilities starting with maximal little
group cases, and consider their successive breakings into smaller
subgroups if necessary in view of the above criterion
(\ref{eq:criterion}).

\section{$G=SU(N)$, $\bfR=\ \sqr85$\,; defining representation $\psi_i$ case}
\label{Sec:NJL_SUn-def}

First consider the simplest case in which the fermion belongs to the 
defining representation $\bfR=\YGo\ $ of $G=SU(N)$;
$\psi_I=\psi_i$. Then, $d:=\dim\YGo=N$ and  the irreducible
decomposition of the symmetric product of $\bfR\times\bfR$ is now
trivial, since $\bfR_p$ is unique: 
\begin{equation}
(\YGo\times\YGo)_\Sym = \YGt\ .  
\end{equation}
So, in this case, the irreducible scalar $\Phi_{\subYGt\,IJ}$ is
identical with the general unconstrained symmetric complex 
$N\times N$ matrix
$\Phi_{IJ}$, so that 
the leading order potential is given by
\begin{align}
V^{\text{leading}}=& \sum_{I=1}^N F(v_I^2\,; M^2_\subYGt), \nn
F(x\,; M^2) :=& M^2 x  - f(x),
\label{eq:function-F}
\end{align}
where $v^2_I$ is the eigenvalues of the $d\times d$ Hermitian matrix
$\Phi^\dagger\Phi$. The point here is that the $d$ eigenvalues $v^2_I$
are all independent and are independently determined by the minimum
condition of the common function $F(x \,; M^2_\subYGt)$. Since the
minimum point $x_0$ is uniquely fixed by $f'(x_0)=M^2_\subYGt$,
we can conclude that 
\begin{equation}
v_I^2 = x_0 \ \ \hbox{for}\ \ \forall I \quad \rightarrow\quad \hbox{$N$
 fermions $\psi_I$ all 
have a degenerate mass-square $x_0$}.
\end{equation}
This common mass-square is, of course, non-vanishing only when 
$G_{\subYGt}=1/M^2_{\subYGt}$ is larger than the critical coupling 
$G_{\text{c}}=16\pi^2/\Lambda^2$.  That is, as far as the dynamical
spontaneous breaking occurs, the subgroup $H$ to which the $G=SU(N)$ is
broken down must be 
such that 
\begin{equation}
\begin{cases}
\hbox{i)}&\ \hbox{$\YGo$ of $SU(N)$ is also an $N$-plet under $H$}; \\
\hbox{ii)}&\ \hbox{$\YGt$ of $SU(N)$ contains an $H$-singlet under $H$-irreducible decomposition}. \\
\end{cases} 
\end{equation}
The first condition alone already {\it excludes the dynamical 
breaking into regular subgroup $H$}! This is because, if $H$ is a regular 
subgroup of $SU(N)$, the defining representation $\YGo$ necessarily splits
into plural $H$-irreducible representations. 
And, the special subgroups $H$ of $G=SU(N)$ satisfying this condition i)
are only $SO(N)$ and $\USp(N)$ (for only even $N$ cases for the latter), 
aside from very special
subgroups like $SO(10)$ for the case of $G=SU(16)$. In any cases, it is
only $SO(N)$ that can also satisfy the second condition ii), since the
symmetric tensor $\Phi_{IJ}$  realizes the common mass
$\langle\Phi_{IJ}\rangle\propto\delta_{IJ}$ for $\psi_I$ but
$\delta_{IJ}$ is an invariant tensor only of $SO(N)$.  

We thus conclude: For $G=SU(N)$ NJL theory with fermion $\psi_I$ in
defining representation $\bfR=\YGo$\,, $SU(N)$ is spontaneously broken to
the special subgroup $H=SO(N)$.
\begin{eqnarray}
G=\hbox{$SU(N)$ NJL} & \text{with} &\{\,\psi_I\,\}\in\YGo: \nonumber\\
G=SU(N) &\rightarrow& H=SO(N) \nonumber\\
\psi_I\ \YGo:\ \ \bf{N} &\rightarrow& \bf{N} \nonumber\\
\Phi_{IJ}\ \YGt:\ \ \frac{\bf{N(N+1)}}{\bf 2} &\rightarrow& \overset{\tr\Phi}{\bf{1}} 
+ \overset{\Phi-\tr\Phi/N} {\bf{(N-1)(N+2)/2}}   
\end{eqnarray} 
in which the $N$-plet fermion $\psi_I$ of $SU(N)$ becomes $N$-plet of
$SO(N)$ and the $N(N+1)/2$ dimensional scalars $\Phi_{IJ}\in\YGt$ splits
into an $SO(N)$ singlet trace part $\tr\Phi= \sum_I\Phi_{II}$ and
traceless symmetric part $\Phi_{IJ}-(1/N)\delta_{IJ}\tr\Phi$ of
dimension $ N(N+1)/2-1 = (N-1)(N+2)/2$; the latter scalars are the
Nambu-Goldstone bosons for this breaking  $SU(N)$ $\rightarrow$
$SO(N)$. Indeed, $\dim SU(N)-\dim SO(N)=(N^2-1)-N(N-1)/2=(N+2)(N-1)/2$.  

Before closing this section, we note an interesting general conclusion 
valid for a special coupling case, which can be drawn 
from this simple example; that is, for the general NJL model with 
fermions of general irreducible representation $\bfR$, we always have 
{\it dynamical breaking into a special subgroup}, if the coupling
constants $G_{\bfR_p}=1/M^2_{\bfR_p}$ for $G$-irreducible channels
$\bfR_p$ are all degenerate (i.e., $\bfR_p$-independent). 
Indeed, in such a case, potential $V$
depends only on the unconstrained scalar $\Phi$ because of the identity
(\ref{eq:11}), so that all the fermions get a common mass just in the
same way as in the simplest model in this section.

\section{$G=SU(N)$, $\bfR=\ $\raise-3.1pt\hbox{$\sqr75$}\kern-8.1pt\raise4.4pt\hbox{$\sqr75$}\,; rank-2 anti-symmetric $\psi_{ij}$ case}
\label{Sec:NJL_SUn-Anti}

Next consider the case where the fermion belongs to the rank-2
anti-symmetric representation $\bfR=\YIshort$, so that the index $I$ now
stands for the anti-symmetric pair
$[ij]$ ($i,j=1,\cdots,N;N\geq2$);
$\psi_I=\psi_{ij}=-\psi_{ji}$. Then the
fermion bilinear scalar $\Phi_{IJ}\sim\psi_I\psi_J$ gives symmetric
product $(\bfR\times\bfR)_{\rm S}$ decomposed into the following  
two irreducible representations $\bfR_p$:
\begin{equation}
\left(\ \YIshort\times\YIshort\ \right)_{\rm S} = \ \YI \ +\  \YII\ .
\end{equation}
Namely, we have two irreducible auxiliary scalar fields in this case:
\begin{equation}
\Phi_{\subYI}{}_{\,ij,kl},
\qquad \Phi_{\subYII}{}_{\,ij,kl}.
\end{equation} 
There are the following six maximal little groups $H$  
of $G=SU(N)$, under which these two $SU(N)$ irreducible 
scalars have $H$-singlet components listed in
Table~\ref{tab:SU-maximal-subgroups}.
\begin{table}[htb]
\begin{center}
\begin{tabular}{clll} \hline
& \hbox{Maximal little group $H$ of $SU(N)$} & \multicolumn{2}{c}{\hbox{$H$-singlet in}}  \\ \hline
1)& \hbox{(Regular) $SU(2)\times SU(N-2)$ \ $(N\geq{ 2})$ case} \phantom{\Big|}& \ \Tspan{.5ex}{0pt}{\YII} & \\[1ex]
2)& \hbox{(Regular) $SU(4)\times SU(N-4)$ \ $(N\geq{ 4})$ case} & &\hspace{-.6em}\YI \\[2ex]
3)& \hbox{(Special) $SO(N)$ \ $(N\geq3)$ case} &\ \YII& \\[1ex]
4)& \hbox{(Special) $\USp(N')$  
( $N'=2\left[\frac{N}2\right], \ N\geq4$ ) case} 
&\ \YII\ \text{and}&\hspace{-.6em}\YI \\[2ex]
5)& \hbox{(Special) $SU(4)\times SU(2)$ case for $N=8$} 
&\ \YII & 
\\[2ex]
6)& 
\hbox{(Special) $SO(10)$ case for $N=16$} &\ \Tspan{0pt}{.5ex}{\YII} &\\\hline
\end{tabular}
\end{center}
\caption{%
Six maximal little groups $H$ of $SU(N)$ for two $SU(N)$
 \YII\ and \YI\ irreducible scalars possessing $H$-singlet.
(Regular) or (Special) in front of each little group name 
denotes the distinction whether it is a regular or special subgroup of
$SU(N)$. The case 4) for $N=3$ reduces to the case 1) (Regular)
$SU(2)\simeq \USp(2)$, so was excluded from there. 
For $N=5$, the case 4) $\USp(4)$ is really the maximal little 
group of $\YGtt$\,, but not of $\YGoooo$\,; for the latter $\YGoooo$\,, 
the maximal little group is the case 2) (Regular) $SU(4)$ which contains
the $\USp(4)$ as a subgroup. 
 } 
\label{tab:SU-maximal-subgroups}
\end{table}

As explained before, we start the analysis of the potential with these
breakings into maximal little groups and  consider the possibility of
successive breakings into further smaller subgroups when necessary. 

First, we consider symmetry breaking of the cases 1), 3), 5), and 6)
since their breakings are caused by the $\YII$ condensation alone, 
so, independent of the coupling
constant $G_{\subYI}=M^{-2}_{\subYI}$. 
As far as the coupling constant $G_{\subYII}=M^{-2}_{\subYII}$ is larger
than its critical coupling, we can compare the potential energies for
those breaking cases with one another irrespectively of the
coupling strength $G_{\subYII}$\,.  From Tables~\ref{table:RSU} 
and \ref{table:SU8-SU16-S}, 
we see that the original fermion ${N(N-1)\over2}$plet $\YGoo$\,,
$\psi_I=\psi_{ij}$, of $G=SU(N)$ is also an $H$-irreducible
${N(N-1)\over2}$plet in the case 3) $H=SO(N)$, and also
in the very special
case 6) of $N=16$, $H=SO(10)$.  The potential for those cases is clearly
given by, for any $N$,
\begin{align}
V_{{SO(N)}}(V) &= \frac{N(N-1)}2 F(V^2\,; M_{\subYII}^2)  
=\frac{N(N-1)}2 \bigl(M_{\subYII}^2 V^2 - f(V^2) \bigr);
\label{eq:SO}
\end{align}
for $N=16$,
\begin{align}
V_{{SO(16)}}(V)=V_{{SO(10)}}(V)
=120 F(V^2\,; M_{\subYII}^2)  
=120\bigl(M_{\subYII}^2 V^2 - f(V^2) \bigr).
\end{align}
Since $V^2$ can be chosen to be the minimum of the function 
$F(V^2\,; M_{\subYII}^2)$, then this potential clearly realizes the
lowest possible value for the 
breakings into this channel scalar $\Phi_{\subYII}$. We can thus forget 
about the other possibilities of 1) and 5), henceforth.

For the other coupling strength cases, $M^2_{\subYI}\leq M^2_{\subYII}$,
we need to consider the condensations into the channel $\Phi_{\subYI}$
also and evaluate the potential in more detail by identifying the
explicit form of the scalar VEVs. So let us now turn to this task.  

\subsection{Scalar VEV and potential for each case} 
\label{Sec:Maximal-subgroups}

Here we identify the explicit form of the scalar VEVs for
the cases 2), 3), 4), and 6) one by one to evaluate the potential in
detail.

2) For the regular breaking case 2) into $H=SU(4)\times SU(N-4)$ 
$(N\geq{4})$, 
the $H$-singlet scalar is contained only in $\Phi_{\subYI}$ and 
the VEV takes the form:
\begin{equation}
\bigl\langle\Phi_{\subYI}{}_{\,ij,kl}\bigr\rangle 
= \frac1{2\cdot2}\,v\, \epsilon_{ijkl56\cdots N},
\label{Eq:RSU4-VEV}
\end{equation}
where $\epsilon_{ijkl56\cdots N}$ is a rank-$N$ totally anti-symmetric
tensor of $SU(N)$ so that it is non-vanishing only when the first four indices
$i,j,k,l$ all take the values 1 to 4 belonging to the $SU(4)$ subgroup. 
This VEV (\ref{Eq:RSU4-VEV}) gives the following form of fermion mass
matrix for the {\bf 6} independent components
$\psi_{i<j}$ ($1\leq i<j\leq4$) $\{\, \psi_{12},\,\psi_{34},\,
\psi_{13},\,\psi_{24},\,\psi_{14},\,\psi_{23}\,\}\in\YIshort$ of $SU(4)$: 
\begin{equation}
\bigl\langle\,\Phi_{\subYI}\,\bigr\rangle=
\bordermatrix{
       &\psi_{12} &\psi_{34}&\psi_{13}&\psi_{24}&\psi_{14}&\psi_{23}\cr
\psi_{12}&        &  v    &       &       &       &       \cr
\psi_{34}&   v    &       &       &       &       &       \cr
\psi_{13}&        &       &       &  -v   &       &       \cr
\psi_{24}&        &       &  -v   &       &       &       \cr
\psi_{14}&        &       &       &       &       &  v    \cr
\psi_{23}&        &       &       &       &  v    &        
}.
\label{Eq:RSU4-VEV-matrix}
\end{equation}
So, in this case of regular breaking into $SU(4)\times SU(N-4)$, only
these six fermions get mass square $v^2$, so the potential is given by
\begin{equation}
V_{{SU(4)\times SU(N-4)}}(v) = 6 F(v^2 ; M_{\subYI}^2) 
=6 \bigl(M_{\subYI}^2 v^2 - f(v^2) \bigr).
\label{eq:RSU4}
\end{equation}

For $N\geq6$, the remaining subgroup $SU(N-4)$ can be further broken 
by the nonvanishing VEV of the scalar field components
$\Phi_{\subYI\,ij,kl}$ and $\Phi_{\subYII\,ij,kl}$ with
$5\leq i,j,k,l\leq N$, keeping the first $SU(4)$ intact. This breaking
again lowers the potential energy since more fermions becomes massive. 
This successive breaking also can be discussed by simply applying our 
present argument for $SU(N)$ to the case $N \rightarrow N-4$.

3) We already know the potential (\ref{eq:SO}) for the third case 3) 
breaking into $H=SO(N)$. For completeness, however, 
we explicitly write the form of the $H$-singlet scalar component in 
$\Phi_{\subYII}$, which is easily guessed to take the form 
\begin{equation}
\bigl\langle\,\Phi_{\subYII}{}_{ij,kl}\,\bigr\rangle= 
\frac1{2\cdot2}V\, \delta_{ij}^{kl},
\label{eq:SOvev}
\end{equation}
where the multi-index Kronecker's delta is defined by 
\begin{equation}
\delta^{i_1i_2\cdots i_n}_{j_1j_2\cdots j_n}
:=n! \delta^{i_1}_{[j_1}\delta^{i_2}_{j_2}\cdots\delta^{i_n}_{j_n]}
:=\sum_{\sigma:\text{permutations}} 
\mathop{\rm sgn}\sigma\,\delta^{i_1}_{j_{\sigma(1)}}\delta^{i_2}_{j_{\sigma(2)}}\cdots\delta^{i_n}_{j_{\sigma(n)}}.
\end{equation}
These deltas are $SU(N)$-invariant tensors if the upper and lower
indices are distinguished as Hermitian conjugate to each other, while,
if such a distinction of upper and lower indices is neglected, then they
are only invariant under $SO(N)$. Thus the VEV (\ref{eq:SOvev}) only
keeps $SO(N)$ while violating the $G=SU(N)$. Under the VEV
(\ref{eq:SOvev}), however, all $N(N-1)/2$ components of fermions
$\psi_{ij}$ get the same mass square $V^2$ and the potential takes the form as 
given in the above Eq.~(\ref{eq:SO}). 

4) The breaking into $\USp(N=2n)$ for even $N=2n$ is most
non-trivial, since both the $G$-irreducible components $\Phi_{\subYI}$
and $\Phi_{\subYII}$ of the scalar $\Phi$ have an $H$-singlet component. 
We should note that $\USp(2n)$ groups have, aside from the usual $SU(N)$ 
invariant tensors $\delta^i_j$ and $\epsilon^{i_1i_2\cdots i_N}$, 
an additional invariant tensor $\Omega^{ij}$, $U^T\Omega U=\Omega$ for 
$\forall U\in{}\USp(2n)$, called symplectic metric whose explicit 
$2n\times2n$ matrix form can be taken to be  
\begin{equation}
\Omega= 
\begin{pmatrix}
 i\sigma_2   &      &       &       \\
        & i\sigma_2  &       &       \\
        &       & \ddots &     \\
        &       &     & i\sigma_2  \\
\end{pmatrix},
\qquad i\sigma_2=
\begin{pmatrix}
        &  1    \\
   -1   &       \\
\end{pmatrix}.
\end{equation}
Then the $H$-singlet component in $\Phi_{\subYI}$ is clearly given by
using the totally anti-symmetric tensor $\epsilon_{i_1i_2\cdots i_N}$ and
the symplectic metric $\Omega^{ij}$ $n-2$ times:
\begin{align}
\bigl\langle\, \Phi_{\subYI}{}_{\,ij,kl} \,\bigr\rangle 
=& 
\frac{v}{2^n(n-2)!} \varepsilon_{ijkl a_1b_1a_2b_2\cdots a_{n-2}b_{n-2}}
\Omega^{a_1b_1}\Omega^{a_2b_2}\cdots\Omega^{a_{n-2}b_{n-2}}.
\end{align}
Note that this VEV for $N=4$, possessing no symplectic metric $\Omega$, 
is $SU(4)$-invariant rather than $\USp(4)$-invariant.

The $H$-singlet component in $\Phi_{\subYII}$ is given by using $\Omega$
twice and by acting the Young symmetrizer ${\cal Y}_{\subYGtt}$ 
to satisfy the required index symmetry:
\begin{align}
\hbox{\Large${\cal Y}$}_{\raisebox{3.5pt}{\Onebox75\raise1pt\llap{{\footnotesize$i$\,}}%
\Onebox75\raise1pt\llap{\footnotesize$k$}}\hskip-14pt%
\raisebox{-3.5pt}{\Onebox75\raise1.2pt\llap{\footnotesize$j$\,}%
\Onebox75\raise.8pt\llap{\footnotesize$l$\,}}}\ 
\Omega^{ij}\Omega^{kl} =&\frac1{16}
\bigl( 1 - (ij)\bigr)\bigl( 1 - (kl)\bigr)
\bigl( 1 + (ik)\bigr)\bigl( 1 + (jl)\bigr)\Omega^{ij}\Omega^{kl} \nn
=& 
\frac14\bigl( 2\Omega^{ij}\Omega^{kl}+\Omega^{ik}\Omega^{jl}-\Omega^{jk}\Omega^{il} \bigr)
\end{align}
with $(ij)$ denoting transposition operator between the indices $i$ and
$j$. So we have 
\begin{align}
\bigl\langle\, \Phi^\dagger_{\subYII}{}^{\,ij,kl} \,\bigr\rangle 
=&
\frac{V}{2\cdot2} \bigl( 2\Omega^{ij}\Omega^{kl}+\Omega^{ik}\Omega^{jl}-\Omega^{jk}\Omega^{il} \bigr).
\end{align} 
With these $H$-singlet VEVs, we can calculate the fermion mass terms by 
a straightforward calculation. 
But, before doing so for general $N=2n$ case, it is helpful to calculate 
these VEV matrices explicitly for the simplest $G=SU(6)$ (i.e., $n=3$)
case. 
Then, among the independent fermions $\psi_I=\psi_{i<j}$, 
we find it convenient to distinguish the 
`diagonal' components $\psi_{2\ell-1,2\ell} (\ell=1,2,\cdots,n)$, 
which appear in the symplectic trace 
$(1/2)\Omega^{ij}\psi_{ij}=\psi_{12}+\psi_{34}+\cdots+\psi_{2n-1,2n}$, from the other 
$2n(n-1)$ `off-diagonal' fermions 
$\psi_{2\ell-1, j}$ or $\psi_{2\ell, j}$ with $j\geq2\ell+1$.  
We put them in the following order explicitly for $n=3$ case:
\begin{equation}
\psi_I = \bigl(
\underbrace{\psi_{12},\ \psi_{34},\ \psi_{56};}
_{\hbox{`diagonal' compts}}\ 
\underbrace{\psi_{13},\,\psi_{24},\,\psi_{14},\,\psi_{23};\ 
\psi_{15},\, \psi_{26},\, \psi_{16},\, \psi_{25};\  
\psi_{35},\, \psi_{46},\, \psi_{36},\, \psi_{45}}_{\hbox{`off-diagonal' compts}}\ 
\bigr). 
\end{equation}
With this independent fermion basis, the $H$-singlet VEV matrices are 
explicitly written as
\begin{align}
\bigl\langle\, \Phi_{\subYI}{}_{\,IJ} \,\bigr\rangle 
=&
\bordermatrix{
       &\psi_{12}&\psi_{34}&\psi_{56}& \psi_{13} &\psi_{24}&\psi_{14}&\psi_{23}&\cdots\cr
\psi_{12}&  0    &  1    &   1   &  &&&&\cr
\psi_{34}&  1    &  0    &   1   &  &&&&\cr
\psi_{56}&  1    &  1    &   0   &  &&&&\cr
\psi_{13}&       &       &       & 0 & -1 & & & \cr
\psi_{24}&       &       &       & -1 & 0 & & & \cr
\psi_{14}&       &       &       &  & & 0 & 1  & \cr
\psi_{23}&       &       &       & & & 1 & 0 & \cr
 \vdots      &       &       &       & & & & &  \ddots \cr
 } \times v, \label{eq:34}
\allowdisplaybreaks[1]\\
\bigl\langle\, \Phi_{\subYII}{}_{\,IJ} \,\bigr\rangle 
=&
\bordermatrix{
       &\psi_{12}&\psi_{34}&\psi_{56}& \psi_{13} &\psi_{24}&\psi_{14}&\psi_{23}&\cdots\cr
\psi_{12}&  3    &  2    &   2   &  &&&&\cr
\psi_{34}&  2    &  3    &   2   &  &&&&\cr
\psi_{56}&  2    &  2    &   3   &  &&&&\cr
\psi_{13}&       &       &       & 0 & 1 & & & \cr
\psi_{24}&       &       &       & 1 & 0 & & & \cr
\psi_{14}&       &       &       &  & & 0 & -1  & \cr
\psi_{23}&       &       &       & & & -1 & 0 & \cr
 \vdots      &       &       &       & & & & &  \ddots \cr
} \times V.
\label{eq:35}
\end{align} 
Note that these matrices are orthogonal to each other, $\tr
\bigl(\langle\Phi_{\subYI}\rangle\langle\Phi_{\subYII}\rangle\bigr)=0$,
as they should be. 
  
Taking these explicit matrix forms into account, we can now write down 
the result for the general $n$ case: 
\begin{align}
(\psi_{ij}\psi_{kl})
\bigl\langle\, \Phi^\dagger_{\subYI}{}^{\,ij,kl} \,\bigr\rangle 
=&
v\Bigl\{ \bigl((\psi_{12}+\psi_{34}+\cdots+\psi_{2n-1,2n})^2 
-(\psi_{12}^2+\psi_{34}^2+\cdots+\psi_{2n-1,2n}^2) \bigr) \nn
&\hspace{1.5em}{}+ 2 \bigl( \sum_{\ell=1}^{n-1}\sum_{m=\ell+1}^n
(-\psi_{2\ell-1,2m-1}\psi_{2\ell,2m}+\psi_{2\ell-1,2m}\psi_{2\ell,2m-1})\bigr) \Bigr\},
\label{eq:36} \\
(\psi_{ij}\psi_{kl}) \bigl\langle\, \Phi^\dagger_{\subYII}{}^{\,ij,kl} \,\bigr\rangle 
=&
V \Bigl\{ \bigl( 2(\psi_{12}+\psi_{34}+\cdots+\psi_{2n-1,2n})^2 
+(\psi_{12}^2+\psi_{34}^2+\cdots+\psi_{2n-1,2n}^2) \bigr) \nn
&\hspace{1.5em}{}- 2 \bigl( \sum_{\ell=1}^{n-1}\sum_{m=\ell+1}^n
(-\psi_{2\ell-1,2m-1}\psi_{2\ell,2m}+\psi_{2\ell-1,2m}\psi_{2\ell,2m-1})\bigr) \Bigr\},
\label{eq:37}
\end{align}
where the first lines of Eqs.~(\ref{eq:36}) and (\ref{eq:37}) 
are for the terms containing only the 
$n$ `diagonal' fermions $\psi_{2\ell-1,2\ell}\, (\ell=1,2,\cdots,n)$, 
and the second lines are 
for the bilinear terms of the other $2n(n-1)$ `off-diagonal' fermions. 
Note that the second lines consist of $n(n-1)$ 
bilinear terms so that all the off-diagonal fermions appear only once
there.  

We can now find the eigenvalues of these matrices  
$\langle\Phi_{\subYI}\rangle$ and
$\langle\Phi_{\subYII}\rangle$. Calculating 
separately the `diagonal' component sector and `off-diagonal' 
component sector, we find the eigenvalues for $SU(2n)$ case
\begin{align}
\langle\Phi_{\subYI}\rangle:\ &  
v\ \Bigl(\  
\overbrace{\ n-1, -1, -1, \cdots, -1}^{\hbox{$n$ `diagonal' compts}};\  
\overbrace{-1,+1, -1,+1; \cdots; -1,+1, -1,+1}^{\hbox{$2n(n-1)$ `off-diagonal' compts}} \ 
\Bigr), \nn
\langle\Phi_{\subYII}\rangle:\ &
V \Bigl( \ 
2n+1, +1, +1, \cdots, +1; \ 
+1,-1, +1,-1; \cdots; +1,-1, +1,-1 \ 
\Bigr). 
\end{align}
Recall that the fermion mass-square eigenvalues are given by the
eigenvalues of $\langle\Phi^\dagger\rangle\langle\Phi\rangle$  with the
total scalar field $\Phi=\Phi_{\subYI}+\Phi_{\subYII}$. 
We, therefore, have the fermion mass-square eigenvalues as
\begin{equation}
\langle\Phi^\dagger\rangle\langle\Phi\rangle:  
\Bigl( \ 
\bigl((2n+1)V+(n-1)v\bigr)^2,\ 
\overbrace{(V-v)^2,\ (V-v)^2,\ \cdots,\ (V-v)^2}^{(n-1)+2n(n-1)=(2n+1)(n-1) \text{times}}\ 
\Bigr).
\end{equation}
Note that this splitting pattern of fermion mass-squared eigenvalues
correctly reflects the decomposition of $SU(2n)$ $\YIshort$ into
$\USp(2n)$-irreducible representations: that is, under
$SU(2n)\supset\USp(2n)$
\begin{equation}
\YIshort:\ \ {\bf \frac{2n(2n-1)}2} = {\bf (2n+1)(n-1)} + {\bf 1}\, ,
\label{eq:SU2USpbranching}
\end{equation}
where the $\USp(2n)$-singlet component is given by 
the symplectic trace $\propto\Omega^{ij}\psi_{ij}$.
Then, noting 
\begin{align}
\tr \langle\Phi^\dagger_{\subYI}\rangle\langle\Phi_{\subYI}\rangle=&
(n-1)^2v^2+ (2n+1)(n-1)v^2= 3n(n-1)v^2, \nn
\tr \langle\Phi^\dagger_{\subYII}\rangle\langle\Phi_{\subYII}\rangle=&
(2n+1)^2V^2+ (2n+1)(n-1)V^2= 3n(2n+1)V^2,
\end{align}
we thus find the potential for this breaking $SU(2n)\rightarrow\USp(2n)$: 
\begin{align}
V_{{\USp(2n)}}=&
M_{\subYI}^2\bigl( 3n(n-1)v^2\bigr) +
M_{\subYII}^2\bigl( 3n(2n+1)V^2 \bigr) \nn
& {}- f\Bigl( \bigl((2n+1)V+(n-1)v\bigr)^2 \Bigr) - (2n+1)(n-1) f\bigl( (V-v)^2 \bigr)
\nn
=&
(M_{\subYII}^2-M_{\subYI}^2)\bigl( 3n(2n+1)V^2 \bigr)  \nn
&{}+F\Bigl( \bigl((2n+1)V+(n-1)v\bigr)^2; M_{\subYI}^2 \Bigr)
+ (2n+1)(n-1) F\bigl( (V-v)^2; M_{\subYI}^2\bigr) \nn
=&
(M_{\subYI}^2-M_{\subYII}^2)\bigl( 3n(n-1)v^2 \bigr) \nn
&{}+F\Bigl( \bigl((2n+1)V+(n-1)v\bigr)^2; M_{\subYII}^2 \Bigr)
+ (2n+1)(n-1) F\bigl( (V-v)^2; M_{\subYII}^2\bigr).
\label{eq:USp}
\end{align}
where the identity (\ref{eq:11}) has been used in going to the second and third 
expressions.

6) Finally, for the case 6) of $SU(16)\to SO(10)$, the potential is the
same as that in Eq.~(\ref{eq:SO}) with $N=16$ for the case 3) of
$SU(16)\to SO(16)$. But the form of the $H$-singlet scalar component in
$\Phi_{\subYII}$ is of course different from the latter case one
(\ref{eq:SOvev}), and is given by  
\begin{equation}
\bigl\langle\, \Phi_{\subYII\, ij,kl} \,\bigr\rangle 
= \frac{V}{2^4 3!}(\sigma_{abc}C)_{ij}(\sigma_{abc}C)_{kl},
\end{equation}
where $\sigma_{abc}=\sigma_{[a}\bar\sigma_b\sigma_{c]}$ of $SO(10)$ Weyl
spinor $\gamma$-matrices with $a,b,c$ being $SO(10)$-vector indices 
and $C$ being the charge conjugation matrix. 
The potential degeneracy between the two breakings $SU(16)\to SO(10)$
and $SU(16)\to SO(16)$ actually causes a very interesting 
mixing phenomenon of
the two vacua, $SO(16)$ and $SO(10)$ ones, which totally breaks 
$SU(16)$ symmetry while keeping the
mass degeneracy of ${\bf 120}$ fermions realizing the lowest
potential value. We explain this phenomenon in Appendix
\ref{sec:su16_so16-so10} in some detail.

\subsection{Which symmetry breaking is chosen?}
\label{Sec:Symmetry-breaking}

Now that the potentials are obtained for the cases 2), 3), 4), and 6),
we can compare them and decide which case realizes the lowest potential
value for various cases of coupling constants. 
Let us discuss three cases, 
(a) $M^2_{\subYI}>M^2_{\subYII}$ $(G_{\subYI}<G_{\subYII})$,
(b) $M^2_{\subYI}=M^2_{\subYII}$ $(G_{\subYI}=G_{\subYII})$, and 
(c) $M^2_{\subYI}<M^2_{\subYII}$ $(G_{\subYI}>G_{\subYII})$, separately.
It is also necessary to discuss even and odd $N(\geq3)$ cases, 
separately, since 
the maximal little group $\USp(N'=2n)$ for the case 4) is also 
a maximal subgroup of $SU(N=2n)$ for even $N$, but not so for odd
$N=2n+1$. In evaluating the potential henceforth, 
we assume that the theory shows the spontaneous symmetry breaking; that
is, the larger coupling constant, at least, is larger than the critical 
coupling constant, ${\rm Min}(G_{\subYGoooo}, G_{\subYGtt}) > G_{\rm cr}$.

\subsection*{\underline{Even $N\geq4$}}

We have already known that for $N=16$ the potentials for the cases 3) and
6) are the same. Here, we need to consider only the potentials for the
cases 2), 3) and 4). 

\subsubsection*{(a) $M^2_{\subYI}>M^2_{\subYII}$ case}

We first compare the potential for 2) $SU(4)\times SU(N-4)$ $(N\geq{4})$
and 3) $SO(N)$ cases. 
\begin{align}
V_{{SU(4)\times SU(N-4)}}(v) =& 6 F(v^2 ; M_{\subYI}^2)
= 6 \bigl(M_{\subYI}^2-M_{\subYII}^2\bigr)v^2 +
6 F(v^2 ; M_{\subYII}^2) 
> 6 F(v^2 ; M_{\subYII}^2) \nn
\geq& 
6 F(V_0^2 ; M_{\subYII}^2) \geq 
\frac{N(N-1)}2 F(V_0^2 ; M_{\subYII}^2) = 
V_{{SO(N)}}\big|_{\text{min}},
\label{Eq:RSU4-SON}
\end{align}
where $V_0^2$ is the minimum point $x=V_0^2$ of the function 
$F(x; M_{\subYII})$ as introduced above. 
Note that $F(V_0^2 ; M_{\subYII}^2)<0$ 
because of the symmetry breaking assumption.
The above inequality holds for $\forall v$.
Therefore, we find for $N\geq{4}$
\begin{equation}
V_{{SU(4)\times SU(N-4)}}\big|_{\text{min}} > 
V_{{SO(N)}}\big|_{\text{min}}.
\end{equation}

Next, we compare the potential for 3) $SO(N)$ and 4) $\USp(N)$ cases.
From Eq.~(\ref{eq:USp})
\begin{align}
V_{{\USp(2n)}}(v,V)=& 
(M_{\subYI}^2-M_{\subYII}^2)\bigl( 3n(n-1)V^2 \bigr)  \nn
&\hspace{-2em}{}+F\Bigl( \bigl( (2n+1)V+(n-1)v\bigr)^2; M_{\subYII}^2 \Bigr)
+ (2n+1)(n-1) F\bigl( (V-v)^2; M_{\subYII}^2\bigr).
\label{eq:USp2}
\end{align}
So, since $(M_{\subYI}^2-M_{\subYII}^2)n(n-1)V^2>0$ in this case, we
have for even $N=2n$, 
\begin{align}
V_{{\USp(2n)}}(v,V)>& 
F\Bigl( \bigl( (2n+1)V+(n-1)v\bigr)^2; M_{\subYII}^2 \Bigr)
+ (2n+1)(n-1) F\bigl( (V-v)^2; M_{\subYII}^2\bigr) \nn
\geq&\bigl(1+ (2n+1)(n-1)\bigr) F\bigl( V_0^2; M_{\subYII}^2\bigr) 
=V_{{SO(N)}}\big|_{\text{min}}.
\label{eq:USp3}
\end{align}
Thus, the $SO(N)$ vacuum realizes the lowest potential value and we can
conclude that the symmetry breaking in this case is also a breaking to
special subgroup:
\begin{equation}
SU(N) \quad \rightarrow\quad SO(N)\ .
\end{equation}

\subsubsection*{(b) $M^2_{\subYI}=M^2_{\subYII}$ case}

We first compare the potential for 2) $SU(4)\times SU(N-4)$ $(N\geq{4})$
and 3) $SO(N)$ cases. From the same discussion 
as in Eq.~(\ref{Eq:RSU4-SON}) for the previous 
(a) $M^2_{\subYI} > M^2_{\subYII}$ case, 
we find for $N\geq4$
\begin{equation}
V_{{SU(4)\times SU(N-4)}}\big|_{\text{min}}{ \geq} 
V_{{SO(N)}}\big|_{\text{min}}, 
\end{equation}
where the equality holds only for $N=4$. 

Next, we compare the potential for 3) $SO(N)$ and 4) $\USp(N)$
cases. This case of degenerate couplings was already discussed generally
at the end of the previous section. We know that all the fermions get a
common mass after symmetry breaking so that the breaking must be down to a
special subgroup. In this case, we have two possibilities for the
special subgroup, $SO(N)$ and $\USp(N=2n)$, which correspond to cases 3)
and 4) breaking, respectively. 
At first sight, the latter $SU(N)\to \USp(N=2n)$ breaking case seems not 
realizing a common mass for all the fermions $\psi_{ij}$ but 
gives two mass square values, since the $N(N-1)/2$-plet
fermion $\psi_{ij}$ splits into a singlet ${\bf1}$ and the rest
${\bf(2n+1)(n-1)}$ under 
$H=\USp(N=2n)$ as already seen in Eq.~(\ref{eq:SU2USpbranching}).   
In the absence of the term  
$(M_{\subYI}^2-M_{\subYII}^2)\bigl( 3n(n-1)V^2 \bigr)$, however, 
$V_{{\USp(2n)}}$ potential (\ref{eq:USp}) takes the form
\begin{eqnarray}
V_{{\USp(2n)}}(v,V)= 
F\bigl( (2n+1)V+(n-1)v\bigr)^2; M_{\subYII}^2 \bigr)
+ (2n+1)(n-1) F\bigl( (V-v)^2; M_{\subYII}^2\bigr). \nonumber
\end{eqnarray}
Since $v$ and $V$ are two independent variables 
corresponding to the VEVs $\bigl\langle\Phi_{\subYI}\bigr\rangle$ and $\bigl\langle\Phi_{\subYII}\bigr\rangle$, respectively, 
the two mass-square parameters $((2n+1)V+(n-1)v)^2$ and $(V-v)^2$ can 
be varied independently so as to choose the minimum $V^2_0$ of the function 
$F( x; M_{\subYII}^2)$. Indeed, two points
\begin{equation}
\begin{cases}
V=V_0/3        &  \\
v=-2V_0/3        & 
\end{cases}
\quad \rightarrow\quad 
\begin{cases}
(2n+1)V+(n-1)v=V_0  &  \\
V-v=V_0        & 
\end{cases}
\label{eq:sol}
\end{equation}
and
\begin{equation}
\begin{cases}
V=(2-n)V_0/3n        &  \\
v=(2n+2)V_0/3n        & 
\end{cases}
\quad \rightarrow\quad 
\begin{cases}
(2n+1)V+(n-1)v=V_0  &  \\
V-v=-V_0        & 
\end{cases}
\label{eq:sol2}
\end{equation}
realize the minimum, and then $1+(2n+1)(n-1)=N(N-1)/2$ fermions all have 
a degenerate mass-square $V_0^2$ also in these $\USp(N=2n)$ vacua. 
(We notice that the latter $\USp(N=2n)$ vacuum (\ref{eq:sol2}) for 
$N=4$ reduces to the $SU(4)\times SU(N-4)=SU(4)$ vacuum realized by 
$\VEV{\Phi_{\subYGtt}\,}=v$ alone, i.e., with $V=0$). 

Recalling the expression 
(\ref{eq:SO}) for the $SO(N)$ potential, 
we see that both $\USp(2n)$ and $SO(N)$ vacua realize the degenerate 
lowest potential minimum in this case:
\begin{equation}
V_{{\USp(N=2n)}}\big|_{\text{min}} =\frac{N(N-1)}2 
F( V_0^2; M_{\subYII}^2)
=V_{{SO(N)}}\big|_{\text{min}},
\label{eq:Vusp2n=Vso2n}
\end{equation}
and we again conclude the breaking into special subgroups also in this
case: 
\begin{equation}
SU(N)\quad \rightarrow\quad SO(N)\ \hbox{or}\ \USp(N=2n)\ ,
\end{equation}
and, for $N=4$ case, in particular, 
\begin{equation}
SU(4)\quad \rightarrow\quad SO(4)\ \hbox{or}\ \USp(4)\ \hbox{or}\ SU(4)\ ,
\end{equation}
although the last SU(4) vacuum breaks no symmetry but is merely a
bilinear fermion condensation.

\subsubsection*{(c) $M^2_{\subYI} < M^2_{\subYII}$ case}

Since the coupling $G_{\subYI}$ becomes stronger in this region, 
we can intuitively 
guess that the $\USp(N=2n)$ vacuum realizes the lower potential value
than the $SO(N)$ one. It can indeed be shown explicitly as follows. 
If we put the above two points (\ref{eq:sol}) and (\ref{eq:sol2}) 
into the expression (\ref{eq:USp}) for 
the potential $V_{{\USp(2n)}}(v,V)$, then, we have
\begin{align}
V_{{\USp(2n)}}\left(-\frac{2V_0}3,\,\frac{V_0}3\right) &=
(M_{\subYI}^2-M_{\subYII}^2) \frac{4n(n-1)}3V_0^2
+V_{{SO(N)}}\big|_{\text{min}}. \nn
V_{{\USp(2n)}}\left(\frac{2(n+1)}{3n}V_0,\,\frac{2-n}{3n}V_0\right) &=
(M_{\subYI}^2-M_{\subYII}^2) \frac{4(n-1)(n+1)^2}{3n}V_0^2
+V_{{SO(N)}}\big|_{\text{min}}.
\label{eq:VUSp-c}
\end{align}
Since the first terms on the RHSs are negative in this case, $\USp(2n)$
potential $V_{{\USp(2n)}}(v,V)$ at these points 
already take values lower than the minimum of the $SO(N)$ potential.
The true minimum of $V_{{\USp(2n)}}(v,V)$ must be lower than these,
implying 
\begin{equation}
V_{{\USp(2n)}}\big|_{\text{min}}
<V_{{SO(N)}}\big|_{\text{min}}.
\end{equation}

So, we next compare the values of $V_{{SU(4)\times SU(N-4)}}$
and $V_{{\USp(2n)}}$ for cases 2) $SU(4)\times SU(N-4)$
$(N\geq4)$ and 4) $\USp(N)$. 

We should first consider a special case $N=2n=4$ (i.e., $n=2$), 
in which $H=SU(4)\times SU(N-4)$ is just $H=SU(4)$ implying no breaking
of $G=SU(4)$. However, the $SU(4)$ vacuum is realized by the
condensation into the channel $\langle\Phi_{\subYGoooo}\,\rangle=v $ and
the potential is given by
$6 F(v^2; M^2_{\subYGoooo}\,)$. All the {\bf 6} components of fermion 
get a common mass square $v_0^2$ realizing the minimum of the function 
$F(x; M^2_{\subYGoooo}\,)$, so it is clear that this $SU(4)$ vacuum
realizes the lowest potential in this coupling region 
$M^2_{\subYGoooo}\,< M^2_{\subYGtt}\,$. (As noted above, the 
second $\USp(4)$ vacuum (\ref{eq:sol2}) for $n=2$ is identical with this
$SU(4)$ vacuum since $V=0$.)  We thus conclude for $N=4$ that 
\begin{equation}
V_{{SU(4)}}\big|_{\text{min}}
<V_{{\USp(4)}}\big|_{\text{min}}
<V_{{SO(4)}}\big|_{\text{min}}.
\label{eq:SU4<USp4}
\end{equation}

Now, we have to consider the general cases $N=2n\geq6$ (i.e., $n\geq3$). 
We here want to show that the opposite to the $N=4$ case holds for this 
general case $N\geq6$; that is, 
$V_{{SU(4)\times SU(N-4)}}\big|_{\text{min}}
>V_{{\USp(N=2n)}}\big|_{\text{min}}$.

{
To show this, we first define the difference as a function of $M^2$
\begin{align}
\Delta(M^2):=&
V_{{\USp(2n)}}\left(v,V\right)\Big|_{\text{min}}
-V_{{SU(4)\times SU(N-4)}}(v)\Big|_{\text{min}},
\label{eq:difference}
\end{align}
and examine its behavior over the region $M_{\subYII}^2>M^2\geq0$, 
where we have simply written $M^2$ to denote $M_{\subYI}^2$ for brevity
and use it for a while hereafter.
We denote $v_0$ as the minimum point of the function 
$F\bigl( x; M^2\bigr)= M^2x -f(x)$ %defined in Eq.~(\ref{eq:function-F})
so that it is a function 
$v_0^2(M^2)$ implicitly determined by
\begin{equation}
M^2=f'(v_0^2). % \ \ \hbox{so that} \ \ 1 = f''(v_0^2) \frac{dv_0^2}{dM^2}.
\label{eq:v0}
\end{equation}
At the boundary $M^2=M^2_{\subYII}$, we already know 
that $\Delta(M^2)$ is negative for $n\geq3$; 
indeed, using the values $V_{{\USp(2n)}}(v,V)\big|_{\text{min}}$ in 
Eq.~(\ref{eq:Vusp2n=Vso2n}) and 
$V_{{SU(4)\times SU(N-4)}}\big|_{\text{min}}=6 F(v_0^2 ; M^2)$
in Eq.~(\ref{eq:RSU4})
%we should use the VEVs $v=2(n+1)v_0/3n, V=(2-n)v_0/3n$
%for $M^2=M^2_{\subYII}$, whose vacuum is smoothly connected with
%$M^2<M^2_{\subYII}$ region;
we have, at $M^2=M^2_{\subYII}$,% and for $n\geq3$,
\begin{align}
 \Delta(M^2=M^2_{\subYII})&\simeq
 V_{{\USp(2n)}}(v, V)
 \Big|_{\text{min}}
-V_{{SU(4)\times SU(N-4)}}(v)\Big|_{\text{min}} \nonumber\\
 &= \bigl(n(2n-1)-6\bigr) F\bigl(v_0^2; M^2\bigr)<0,
 \label{eq:bdvalue}
\end{align}
since $n(2n-1)-6\geq9$ for $n\geq3$ and $F\bigl(v_0^2; M^2\bigr)<0$. 
%is where $V_{{SU(4)\times SU(N-4)}}\big|_{\text{min}}=6 F(v_0^2 ; M^2)$
%from Eq.~(\ref{eq:RSU4}).
We will show that $d\Delta(M^2)/dM^2\geq0$ in the present 
region $M^2_{\subYII}\geq M^2\geq0$. 
Then, if we see $\Delta(M^2)$ in the region 
$M^2_{\subYII}\geq M^2\geq0$ from $M^2=M^2_{\subYII}$  
toward the direction of $M^2$ going to smaller to zero 
(the direction of the coupling constant $G_{\subYI}=M^{-2}$ going to 
stronger to $\infty$), it  {\it decreases} monotonically from the 
initial negative value Eq.~(\ref{eq:bdvalue}) at $M^2=M^2_{\subYII}$, 
implying that it is always negative 
in $M^2_{\subYII}\geq M^2\geq0$.

The derivative of $\Delta(M^2)$ 
with respect to $M^2$ is evaluated as
\begin{align}
\frac{d}{dM^2}\Delta(M^2)=&\frac{d}{dM^2}
\left(V_{{\USp(2n)}}(v,V)\Big|_{\text{min}}
 -V_{{SU(4)\times SU(N-4)}}(v)\Big|_{\text{min}}\right)\nonumber\\
=&\frac{\partial}{\partial M^2}
\Bigl(V_{{\USp(2n)}}(\bar{v}, \bar{V})%\Big|_{\text{min}}
 -V_{{SU(4)\times SU(N-4)}}(v_0)%\Big|_{\text{min}}
 \Bigr)
 +\frac{\partial V_{{\USp(2n)}}}{\partial v}\Big|_{\text{min}}
 \frac{\partial\bar{v}}{\partial M^2} +\cdots
 \nonumber\\
%&\hspace{1em}
% +\left(\frac{\partial V_{{\USp(2n)}}}{\partial v}\Big|_{\text{min}}
% \frac{\partial\bar{v}}{\partial M^2}
% -\frac{\partial V_{{\USp(2n)}}}{\partial V}\Big|_{\text{min}}
% \frac{\partial\bar{V}}{\partial M^2}
% \right)
% +\frac{\partial V_{{SU(4)\times SU(N-4)}}}{\partial v}\Big|_{\text{min}}
% \frac{\partial v_0}{\partial M^2}
%\nonumber\\
%=&\frac{\partial}{\partial M^2}
%\left(V_{{\USp(2n)}}(v,V)\Big|_{\text{min}}
% -V_{{SU(4)\times SU(N-4)}}(v)\Big|_{\text{min}}\right)\nonumber\\
=&3n(n-1)\bar{v}^2-6v_0^2,
\label{eq:difference-USp2n-SU4}
\end{align}
where $(\bar{v}, \bar{V})$ is the value of $(v, V)$ at the minimum 
point of $V_{{\USp(2n)}}(v,V)$, and the explicit $M^2$-dependence 
has been found in the expressions 
(\ref{eq:USp}) for $V_{{\USp(2n)}}(\bar{v},\bar{V})$
and (\ref{eq:RSU4}) for $V_{{SU(4)\times SU(N-4)}}(v_0)$. 
Note that the implicit $M^2$-dependence here through 
$\bar{v}(M^2), \bar{V}(M^2)$ and $v_0(M^2)$ does not contribute 
because of the stationarity of the potential at the minimum: 
\begin{align}
 \frac{\partial V_{{\USp(2n)}}}{\partial v}\Big|_{\text{min}}
=\frac{\partial V_{{\USp(2n)}}}{\partial V}\Big|_{\text{min}}
=\frac{\partial V_{{SU(4)\times SU(N-4)}}}{\partial v}\Big|_{\text{min}}
=0.
\label{eq:stationary-condition-USp2n-SU4}
\end{align}
The minimum point $(\bar{v}, \bar{V})$ of $V_{{\USp(2n)}}(v,V)$ 
is found by the first and second equations in
Eq.~(\ref{eq:stationary-condition-USp2n-SU4}) by using Eq.~(\ref{eq:USp}):
\begin{align}
&3nM^2\bar{v}=f'(v_1^2)v_1+(2n+1)f'(v_2^2)v_2,\nonumber\\
&3nM_{\subYII}^2(-\bar{V})=-f'(v_1^2)v_1+(n-1)f'(v_2^2)v_2,
\label{eq:stationary-condition-USp2n-dM^2} 
\end{align}
where $v_1:=(2n+1)\bar{V}+(n-1)\bar{v}$ and $v_2:=\bar{v}-\bar{V}$ are 
(square root of) the arguments of the two $f$ functions in Eq.~(\ref{eq:USp}) 
at the minimum 
point. Inserting the inverse relation 
\begin{align}
3n\bar{v}=v_1+(2n+1)v_2, \quad 
3n(-\bar{V})=-v_1+(n-1)v_2, 
\end{align}
Eq.~(\ref{eq:stationary-condition-USp2n-dM^2}) can be rewritten into
%We will show the positivity of $d\Delta(M^2)/dM^2$ for 
%$M^2<M^2_{\subYII}$ region by using Eq.~(\ref{eq:difference-USp2n-SU4}).
%To do so, we need to find the relations between the VEVs $v_j (j=0,1,2)$.
%From the stationary conditions in
%Eq.~(\ref{eq:stationary-condition-USp2n-dM^2}), we find 
\begin{align}
\big(f'(v_1^2)-M^2\big)v_1+(2n+1)\big(f'(v_2^2)-M^2\big)v_2&=0,
\label{eq:B-v12-1}\\
-\big(f'(v_1^2)-M_{\subYII}^2\big)v_1
+(n-1)\big(f'(v_2^2)-M_{\subYII}^2\big)v_2&=0.
\label{eq:B-v12-2}
\end{align}
%where we define $B_i, B_i^{\subYII} (i=1,2)$ as
%\begin{align}
%B_i:=f'(v_i^2)-M^2,\ \ \
%B_i^{\subYII}:=f'(v_i^2)-M_{\subYII}^2.
%\label{eq:Bs} 
%\end{align}
In order for this simultaneous Eqs.~(\ref{eq:B-v12-1}) and (\ref{eq:B-v12-2}) 
to have non-vanishing solution, 
\begin{align}
\det\left(
\begin{array}{cc}
 f'(v_1^2)-M^2&(2n+1)\big(f'(v_2^2)-M^2\big)\\
-\big(f'(v_1^2)-M_{\subYII}^2\big) & (n-1)\big(f'(v_2^2)-M_{\subYII}^2\big)\\
\end{array}
\right) 
%=(n-1)B_1B_2^{\subYII}+(2n+1)B_1^{\subYII}B_2=0.
\end{align}
must vanish, so that we obtain
\begin{align}
 f'(v_1^2)-M^2 % B_1
% =-\left(\frac{2n+1}{n-1}\right)
% \frac{B_1^{\subYII}}{B_2^{\subYII}}B_2
 =-\left(\frac{2n+1}{n-1}\right)\alpha\cdot \big(f'(v_2^2)-M^2\big),
\label{eq:relation-B1-B2} 
\end{align}
where we have defined a parameter
\begin{align}
 \alpha:= %\frac{B_1^{\subYII}}{B_2^{\subYII}}
 \frac{M_{\subYII}^2-f'(v_1^2)}{M_{\subYII}^2-f'(v_2^2)}.
\label{eq:alpha-def} 
\end{align}
From Eq.~(\ref{eq:B-v12-2}), we also have
\begin{align}
 v_1
%=(n-1)
% \frac{M_{\subYII}^2-f'(v_1^2)}{M_{\subYII}^2-f'(v_2^2)}v_2
 =(n-1)\alpha^{-1}v_2.
\label{eq:relation-v1-v2} 
\end{align}

From these equations, we can now discuss the size ordering among $v_1^2, v_2^2$ 
and $v_0^2$. 
If the coupling $G_{\subYII}$ is moved below the critical value 
$G_{\text{cr}}$, i.e., 
$M_{\subYII}^2>f'(0)$, while keeping $G_{\subYI}>G_{\text{cr}}$, then 
the parameter $\alpha$ in Eq.~(\ref{eq:alpha-def}) is clearly positive. 
So we henceforth consider only the solution $(v_1, v_2)$ of 
Eqs.~(\ref{eq:B-v12-1}) and (\ref{eq:B-v12-2}) which satisfies 
$\alpha>0$.\footnote{%
When both coupling constants $G_{\subYI}$ and $G_{\subYII}$ are above 
critical, there are actually two solutions to the simultaneous 
Eqs.~(\ref{eq:B-v12-1}) and (\ref{eq:B-v12-2}): 
One realizes $\alpha>0$ and reduces to the solution Eq.~(\ref{eq:sol2}) 
in the limit $M^2 \rightarrow M_{\subYII}^2$, and the other realizes 
$\alpha<0$ and reduces to the solution Eq.~(\ref{eq:sol}). However, one can 
convince oneself 
that the latter solution with $\alpha<0$ has the size-ordering 
$v_1^2<v_{\,\subYII}^2<v_2^2<v_0^2$ $\big(f'(v_{\,\subYII}^2):=M_{\subYII}^2\big)$ 
and realizes higher potential value than that realized by the former solution 
with $\alpha>0$ discussed here. In any case, it is enough to prove 
$V_{\USp}|_{\text{min}}<V_{{SU(4)\times SU(N-4)}}|_{\text{min}}$ for 
one solution for the present purpose.}
Then, 
Eqs.~(\ref{eq:relation-B1-B2}), (\ref{eq:alpha-def}) and 
$M^2< M^2_{\subYII}$ tell us that 
either i) $f'(v_1^2)<M^2<f'(v_2^2)<M^2_{\subYII}$ with $\alpha>1$
or ii) $f'(v_2^2)<M^2<f'(v_1^2)<M^2_{\subYII}$ with $1>\alpha>0$
holds, which corresponds to either 
i) $v_1^2>v_0^2>v_2^2$ or ii) $v_2^2>v_0^2>v_1^2$, respectively, 
since $f'(x)$ is a monotonically decreasing 
function and $M^2=f'(v_0^2)$. 
%from Figure~\ref{fig:fx-functions}, for $\alpha>1$, then
%(Since the over-all sign of $v_1, v_2$ is irrelevant to the solution, we can 
%take $v_2>0$ as our convention.) %, then both $v_1$ and $v_2$ are positive by 
However, the case ii) is inconsistent with Eq.~(\ref{eq:relation-v1-v2}), 
which says $v_1^2>v_2^2$ since $(n-1)\alpha^{-1}>1$ for $\alpha<1, n\geq2$. 
Thus we have only the case i), which is consistent with 
Eq.~(\ref{eq:relation-v1-v2}) if $n-1>\alpha>1$.

%Next, Eqs.~(\ref{eq:B-v12-1}) and (\ref{eq:B-v12-2}) lead to 
%\begin{align}
% f'(v_1^2)v_1+f'(v_2^2)v_2(2n+1)&
% =M^2\left(v_1+(2n+1)v_2\right),\\
% -f'(v_1^2)v_1+f'(v_2^2)v_2(n-1)&
% =M_{\subYII}^2\left(-v_1+(n-1)v_2\right).
%\end{align}
%
Now to prove the positivity of Eq.~(\ref{eq:difference-USp2n-SU4}), we need 
an inequality. Recall that $f'(x)$ is a monotonically decreasing 
downward-convex function, so it satisfies the following inequality for
${}^\forall\lambda\in[0,1]$, 
\begin{align}
 f'\left(\lambda x_1+(1-\lambda)x_2\right)\leq 
 \lambda\cdot f'(x_1)+(1-\lambda)f'(x_2).
\label{eq:downward-convex-function} 
\end{align}
Noting the ordering $v_2^2<v_0^2<v_1^2$, we take  
$x_1=v_1^2, \ x_2=v_2^2$ and $\lambda=(v_0^2-v_2^2)/(v_1^2-v_2^2)$,
%$v_2^2<v_0^2<v_1^2$, $\lambda=(v_0^2-v_2^2)/(v_1^2-v_2^2)$, we find
this leads to
\begin{align}
f'(v_0^2)\leq 
\frac{(v_0^2-v_2^2)f'(v_1^2)+(v_1^2-v_0^2)f'(v_2^2)}{v_1^2-v_2^2}.
\end{align}
%where we used $f'(v_0^2)=M^2$. 
Multiplying this by $v_1^2-v_2^2>0$ and inserting 
Eq.~(\ref{eq:relation-B1-B2}) with $M^2=f'(v_0^2)$ there, and dividing it 
with the positive factor $f'(v_2^2)-M^2>0$, we find
\begin{align}
 v_2^2\left(\frac{2n+1}{n-1}\right)\alpha+v_1^2
 \geq v_0^2\left(\frac{2n+1}{n-1}\alpha+1\right).
\end{align}
Further, inserting Eq.~(\ref{eq:relation-v1-v2}), $v_1=(n-1)\alpha^{-1}v_2$, 
we finally find
\begin{align}
 v_2^2\left(\frac{2n+1}{n-1}\alpha+(n-1)^2\alpha^{-2}\right)
 \geq v_0^2\left(\frac{2n+1}{n-1}\alpha+1\right).
\label{eq:relation-v2-v0} 
\end{align}

Now, we can evaluate $d\Delta(M^2)/dM^2$ in
Eq.~(\ref{eq:difference-USp2n-SU4}); the first term is given by
\begin{align}
3n(n-1)\bar{v}^2
 &=\frac{n-1}{3n}
 \left(v_1+(2n+1)v_2\right)^2
% =\frac{n-1}{3n}\left((n-1)\alpha^{-1}+(2n+1)\right)v_2^2\nonumber\\
 \geq 
 %\frac{(n-1)^3}{3n}\left(\alpha^{-1}+\frac{2n+1}{n-1}\right)^2
% \frac{\left(\alpha^{-2}+\frac{2n+1}{n-1}\right)v_0^2}
% {(n-1)^2\left(\alpha^{-3}+\frac{2n+1}{(n-1)^3}\right)}
% =
 \frac{n-1}{3n}\frac{\left(\alpha^{-1}+\frac{2n+1}{n-1}\right)^3}
 {\alpha^{-3}+\frac{2n+1}{(n-1)^3}}v_0^2=:\frac{n-1}{3n}G(\alpha^{-1})v_0^2,
\end{align}
where we have used Eqs.~(\ref{eq:relation-v1-v2}) and
(\ref{eq:relation-v2-v0}). 
An elementary analysis for the function $G(\alpha^{-1})$ over the 
region $n-1>\alpha>1$, i.e., $(n-1)^{-1}<\alpha^{-1}<1$, shows that 
$G(\alpha^{-1})$ is maximum at the starting point $\alpha^{-1}=(n-1)^{-1}$ 
and is a monotonically decreasing function in this region. So    
the minimum of the function $G(\alpha^{-1})$ is
located at $\alpha^{-1}=1$ for $n\geq3; G(1)=27n^2/(n^2-3n+5)$. 
Therefore, we have 
\begin{align}
\frac{d\Delta(M^2)}{dM^2}
&= 3n(n-1)\bar{v}^2-6v_0^2
=\frac{n-1}{3n}\frac{27n^2}{n^2-3n+5}v_0^2-6v_0^2 \nn
&=\frac{9n(n-1)-6(n^2-3n+5)}{n^2-3n+5}v_0^2
=
\frac{3(n-2)(n+5)}{n(n-3)+5}v_0^2\geq0 \quad (\hbox{for }\ n\geq3).
\end{align}
Together with the boundary value $\Delta(M^2_{\subYII})<0$ in 
Eq.~(\ref{eq:bdvalue}), this positivity proves that $\Delta(M^2)$ is negative 
definite in the region $M^2_{\subYII}\geq M^2\geq0$ and hence we can conclude that,
for $N=2n\geq6$, 
\begin{equation}
V_{{\USp(2n)}}\big|_{\text{min}}<
V_{{SU(4)\times SU(N-4)}}\big|_{\text{min}}\,.
\label{eq:USp<SU}
\end{equation}
We thus again conclude the breaking into special subgroups also in this
case $N\geq6$: 
\begin{equation}
SU(N) \quad \rightarrow\quad \USp(N=2n) \,. 
\end{equation}
}
The $SU(2n)$ phase diagrams are shown in the coupling constant 
plane 
$(G_{\subYII}\,,\ G_{\subYI})$ for $N=2$ and $n\geq3\ \not=8$ and
 $n=8$ cases in Figure~\ref{figure:SU2n-phase-diagram}.

\begin{figure}[htb]
\begin{center}
\includegraphics[bb=0 0 33 32, width=50mm]{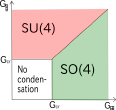}
\hspace{.5em}
\includegraphics[bb=0 0 33 32, width=50mm]{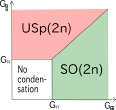}
\hspace{.5em}
\includegraphics[bb=0 0 33 32, width=50mm]{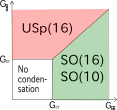}
\end{center}
 \caption{Even $N=2n\geq4$:
 $SU(2n)$ phase diagrams are shown in the coupling
 constant plane 
$(G_{\subYII}\,,\ G_{\subYI})$ for $n=2$ (left figure) and 
$n\geq3,\ \not=8$ (middle figure) and  $n=8$ (right figure) cases. }
\label{figure:SU2n-phase-diagram}
\end{figure}

Here, however, we should comment on the possibility of further breaking
of the $SU(N-4)$ part of $SU(4)\times SU(N-4)$, which exists for $n\geq3$ and 
can actually make
the potential lower as remarked before. However, in this coupling
region, we now know that the breaking $SU(N-4)\to \USp(N-4)$ realizes the
lowest potential energy, so we should consider the possibility of the
successive breaking
$SU(N)\to SU(4)\times SU(N-4)\to SU(4)\times\USp(N-4)$. But, since the  
first and second breakings have no interference, we have   
\begin{eqnarray}
V_{{SU(4)\times\USp(N-4)}}\big|_{\text{min}}
&=& 
V_{{SU(4)\times SU(N-4)}}\big|_{\text{min}}+ 
V_{{\USp(N-4)}}\big|_{\text{min}}.
\end{eqnarray} 
This should be compared with the value
$V_{{\USp(N=2n)}}\big|_{\text{min}}$. 
Although we do not show an explicit proof here, it is almost evident that 
\begin{equation}
V_{{\USp(2n)}}\big|_{\text{min}}<
V_{{SU(4)\times\USp(N-4)}}\big|_{\text{min}}.
\end{equation}
This is because the number of the massive fermions on the $\USp(2n)$
vacuum is much larger than that on the $SU(4)\times\USp(N-4)$ vacuum;
the difference is  
\begin{equation}
n(2n-1) - \left( 6 + (n-2)\bigl(2(n-2)-1\bigr) \right)
= 8n-16 
\end{equation}
which is larger than 8 already at the lowest value $n=3$ here. So the
above conclusion of the $SU(2n)\to \USp(2n)$ breaking is still valid even 
if the possibility of the breaking into non-maximal little groups is
taken into account.

\subsection*{\underline{Odd $N\geq3$}}

From Table~\ref{tab:SU-maximal-subgroups}, for $N=3$, only the case 3)
is possible; for $N\geq5$, the cases 2), 3) and 4) are possible.

Obviously, for $N=3$, $SU(3)$ is broken to the maximal special subgroup
$SO(3)$ as far as $G_{\subYII}$ is larger than its critical coupling.
We will discuss the potentials for $N\geq5$ in detail.

\subsubsection*{(a) $M^2_{\subYI}>M^2_{\subYII}$ and (b) $M^2_{\subYI}=M^2_{\subYII}$ cases}

We first compare the potentials for cases 2) $SU(4)\times SU(N-4)$
$(N\geq5)$ and 3) $SO(N=2n+1)$. The inequality in
Eq.~(\ref{Eq:RSU4-SON}) holds also for odd
$N\geq5$. Therefore, we find for $N=2n+1\geq5$ 
\begin{equation}
V_{{SU(4)\times SU(N-4)}}\big|_{\text{min}} > 
V_{{SO(N=2n+1)}}\big|_{\text{min}}.
\end{equation}

Next, we compare the potentials for cases 3) $SO(N=2n+1)$ and 4)
$\USp(N'=2n)$ $(N\geq5)$. From Eq.~(\ref{eq:USp3}) for (a)
$M^2_{\subYI}>M^2_{\subYII}$ case and  
Eq.~(\ref{eq:Vusp2n=Vso2n}) for (b) $M^2_{\subYI}=M^2_{\subYII}$ case, we know 
\begin{align}
V_{{\USp(2n)}}(v,V)\big|_{\text{min}}\geq& \frac{N'(N'-1)}2 
F( V_0^2; M_{\subYII}^2)=
V_{{SO(2n)}}\big|_{\text{min}}
\end{align}
with equality for the case (b). But, since 
Eq.~(\ref{eq:SO}) tells us the inequality
\begin{equation}
V_{{SO(2n)}}\big|_{\text{min}}
>V_{{SO(2n+1)}}\big|_{\text{min}},
\end{equation}
we have anyway
\begin{equation}
V_{{\USp(N'=2n)}}(v,V)\big|_{\text{min}}
>V_{{SO(N=2n+1)}}\big|_{\text{min}}.
\end{equation}
Thus, the $SO(N=2n+1)$ $(N\geq5)$ vacuum realizes the lowest potential
value and we can conclude that the symmetry breaking in  
these cases $M^2_{\subYI}{ \geq}M^2_{\subYII}$ is 
also a breaking to special subgroup:
\begin{equation}
SU(N=2n+1) \quad \rightarrow\quad SO(N=2n+1)\ .
\end{equation}

\subsubsection*{(c) $M^2_{\subYI}<M^2_{\subYII}$ case}

In this coupling region, the condensation into $\Phi_{\subYGoooo}$ is preferred to 
into $\Phi_{\subYGtt}$. 
Here we first compare the potentials for 2) $SU(4)\times SU(N-4)$
 and 4) $\USp(N'=N-1)$ for $(N\geq5)$. 
The same discussion as in even $N$, given from Eq.~(\ref{eq:difference})
to Eq.~(\ref{eq:USp<SU}), holds if $n\geq3$, so that  
we have, for $N=2n+1\geq7$, 
\begin{eqnarray}
V_{{\USp(N'=2n)}}|_{\text{min}}
 <V_{{SU(4)\times SU(N-4)}}|_{\text{min}}.
\end{eqnarray}
For $N=5$, however, $\USp(N'=4)$ is not the maximal little group of
$\YGoooo$, for which the $SU(4)\times SU(N-4)=SU(4)$ is the maximal
little group. Since six fermions of $SU(4)$ {\bf 6} all can get a common
mass square realizing the minimum of the potential
$F( v^2; M_{\subYI}^2)$ for the $SU(4)$ vacuum case, while they must
split into ${\bf 5}+{\bf1}$ under the subgroup $\USp(4)\subset SU(4)$  
so leading necessarily to the higher energy than the $SU(4)$ case,
\begin{equation}
V_{{SU(4)}}\big|_{\text{min}}
<V_{{\USp(4)}}\big|_{\text{min}}.
\label{eq:SU4<USp4-SU5}
\end{equation}
This is the same inequality as the first part of Eq.~(\ref{eq:SU4<USp4}).

\begin{figure}[htb]
\begin{center}  
\includegraphics[bb=0 0 34 31, width=50mm]{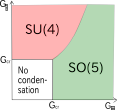}
\hspace{2em}
\includegraphics[bb=0 0 34 31, width=50mm]{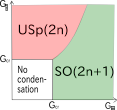}
\end{center}
 \caption{Odd $N=2n+1\geq5$:
 $SU(2n+1)$ phase diagrams (rough sketch) are shown in the
 coupling  constant $(G_{\subYII}, G_{\subYI}\,)$ plane for $n=2$
 (left figure) and  $n\geq3$ (right figure) cases. }
\label{figure:SU2n+1-phase-diagram}
\end{figure}

Next, we compare the potentials for 3) $SO(N)$ and 2)
$SU(4)\times SU(N-4)$  for $N=5$; 4) $\USp(N'=N-1)$ for $N\geq7$.
If $M_{\subYI}^2$ becomes much smaller than $M_{\subYII}^2$\,, 
i.e., the coupling $G_{\subYI}$ becomes much stronger than
$G_{\subYII}$\,, 
then the minimum value $F(v_0^2; M_{\subYI})$ becomes much lower than 
the minimum value $F(V_0^2; M_{\subYII})$. 
The minimum value of the $SU(4)\times SU(N-4)$ potential 
$V_{{SU(4)\times SU(N-4)}}|_{\text{min}}=6 F( v_0^2; M_{\subYI}^2)$
or the $\USp(N'=N-1)$ potential $V_{{\USp(N'=N-1)}}|_{\text{min}}$, for
which the number of the massive fermions is smaller than that for the
$SO(N)$ case, can become lower than the minimum value of the $SO(N)$
potential $V_{{SO(N)}}|_{\text{min}}=(N(N-1)/2)F(V_0^2; M_{\subYII})$ 
for $N\geq5$. Thus, we conclude that, for odd $N\geq5$, the symmetry
breaking pattern depends on whether $M_{\subYI}^2$ is larger or smaller
than a certain value $M_0^2$ which depends on $N$:
for $N=5$
\begin{equation}
SU(5) \quad \rightarrow\quad 
\begin{cases} 
SO(5) & \hbox{for } \exists M_0^2\leq M_{\subYI}^2\leq M_{\subYII}^2 \\
SU(4) & \hbox{for } M_{\subYI}^2\leq\exists M_0^2<M^2_{\subYII} 
\end{cases};
\end{equation} 
for $N\geq7$,
\begin{equation}
SU(N=2n+1) \quad \rightarrow\quad 
\begin{cases} 
SO(N=2n+1) & \hbox{for } \exists M_0^2\leq M_{\subYI}^2\leq M_{\subYII}^2 \\
\USp(N'=2n) & \hbox{for } M_{\subYI}^2\leq\exists M_0^2<M^2_{\subYII} 
\end{cases}.
\end{equation} 
The $SU(2n+1)$ phase diagrams are shown in the coupling constant 
$(G_{\subYII}, G_{\subYI}\,)$ plane for $n=2$ and $n\geq3$ cases, 
in Figure~\ref{figure:SU2n+1-phase-diagram}.

\begin{table}[htb]
\caption{Branching rules of 
$SU(n)\supset SU(m=n-\ell)\times SU(\ell)\times U(1)$, 
$\ {}\supset SO(n)$, and $\ {}\supset\USp(2n)$.}
\label{table:RSU}
\begin{center}
\begin{tabular}{rrcl} \hline\hline 
 & $SU(n)$& $\supset$ & $SU(n-\ell=m)\times SU(\ell)\times U(1)$ \\ \hline
\phantom{$\Big|$}$\Onebox75$\ :& {\bf n} &= & $(\Onebox75\,,\, {\bf1})(\ell)\oplus({\bf1},\, 
\Onebox75)(-m)$ \\
$\YGt$\ :& $\displaystyle\frac{\bf n(n+1)}{\bf2}$ &=& 
 $\left(\YGt\,,\, {\bf1}\right)(2\ell)
 \oplus\left(\Onebox75\,,\, \Onebox75\,\right)(\ell-m)
 \oplus\left({\bf1},\, \YGt\,\right)(-2m)$ \\[1.5ex] \hline
$\YGoo$\ :&  
\phantom{$\Bigg|$}$\displaystyle\frac{{\bf n(n-1)}}{\bf2}$ & =&
 $\left(\Onebox75\,,\, \Onebox75\,\right)(\ell-m)
 \oplus\left(\YGoo\,,\, {\bf 1}\right)(2\ell)
 \oplus\left({\bf 1},\, \YGoo\, \right)(-2m) $ \\[1ex]
$\YGtt$\ :& $\displaystyle\frac{\bf n^2(n+1)(n-1)}{\bf12}$& =& 
 $\left(\YGt\,,\, \YGt\,\right)(2\ell-2m)
 \oplus\left( \YGto\,,\, \Onebox75\,\right)(3\ell-m)$ \\[1ex]
 &&&
 $\oplus\left(\YGtt\,,\, {\bf1}\right)(4\ell)
 \oplus\left(\Onebox75\,,\, \YGto\,\right)(\ell-3m) $ \\[1.5ex]
 &&&
 $\oplus\left(\YGoo\,,\, \YGoo\,\right)(2\ell-2m)
 \oplus\left({\bf 1},\, \YGtt\,\right)(-4m) $ \\[1.5ex]
$\YGoooo$\ :& $\displaystyle\frac{\bf n(n-1)(n-2)(n-3)}{\bf24}$ &=& 
 $\left( \Onebox75\,,\, \YGooo\, \right)(\ell-3m)
 \oplus\left(\YGoo\,,\, \YGoo\, \right)(2\ell-2m)$ \\
 &&& $\oplus\left(\YGooo\,,\, \Onebox75\, \right)(3\ell-m) 
 \oplus\left(\YGoooo\,,\, {\bf1}\right)(4\ell)
 \oplus\left({\bf 1},\, \YGoooo\, \right)(-4m) $ \\[2.5ex] \hline
&$\dim \YGto = \Tspan{.3ex}{.5ex}{\displaystyle\frac{\bf n(n^2-1)}{\bf3}}$ &&
$\dim \YGooo= \displaystyle\frac{\bf n(n-1)(n-2)}{\bf6}$\ \ for $SU(n)$  \\ \hline
\hline
 & $SU(n)$ & $\supset$ & $SO(n)$ \\ \hline
$\Onebox75$\ : & {\bf n} &=& {\bf n} \\
$\YGt$\ : &
${\bf \displaystyle\frac{n(n+1)}2}$ &=& ${\bf \displaystyle\frac{(n-1)(n+2)}2}\oplus{\bf 1}$ \\[1.5ex] \hline
$\YGoo$\ : &
\phantom{$\Bigg|$} ${\bf \displaystyle\frac{n(n-1)}2}$ &=& ${\bf \displaystyle\frac{n(n-1)}2}$ \\
$\YGtt$\ : &
 ${\bf \displaystyle\frac{n^2(n+1)(n-1)}{12}}$ &=&
${\bf \displaystyle\frac{n(n+1)(n+2)(n-3)}{12}}\oplus{\bf \displaystyle\frac{{(n-1)}(n+2)}2}\oplus{\bf 1}$ \\[1ex]
$\YGoooo$\ : &
${\bf \displaystyle\frac{n(n-1)(n-2)(n-3)}{24}}$ &=& ${\bf \displaystyle\frac{n(n-1)(n-2)(n-3)}{24}}$.\\[2ex] \hline
\hline
 & $SU(2n)$ & $\supset$ & $\USp(2n)$ \\ \hline
$\Onebox75$\ : &
{\bf 2n} &=& {\bf 2n} \\
$\YGt$\ : &
${\bf n(2n+1)}$ &=& ${\bf n(2n+1)}$ \\[.5ex] \hline
$\YGoo$\ : &
\phantom{$\bigg|$}${\bf n(2n-1)}$ &=& ${\bf (n-1)(2n+1)}\oplus{\bf 1}$ \\[.5ex]
$\YGtt$\ : &
${\bf \displaystyle\frac{n^2(2n+1)(2n-1)}{3}}$ &=& 
${\bf \displaystyle\frac{n(n-1)(2n-1)(2n+3)}{3}}\oplus{\bf (n-1)(2n+1)}\oplus{\bf 1}$ \\[1.5ex]
$ \YGoooo$\ : &
${\bf \displaystyle\frac{n(n-1)(2n-1)(2n-3)}{6}}$ &=&
${\bf \displaystyle\frac{n(n-3)(2n+1)(2n-1)}{6}}
\oplus{\bf (n-1)(2n+1)}\oplus{\bf 1}$ \\[2ex] \hline
\end{tabular}
\end{center}
\end{table}

\begin{table}[htb]
\caption{Branching rules of 
$SU(8)\supset SU(4)\times SU(2)$ and $SU(16)\supset SO(10)$.
}
\label{table:SU8-SU16-S}
\begin{center}
\begin{tabular}{rrcl} \hline\hline 
 & $SU(8)$ & $\supset$ & $SU(4)\times SU(2)$ \\ \hline
\Tspan{1ex}{0ex}{$\Onebox75$}\ :& {\bf 8} &= & $({\bf 4,2})$\\[0.5em]
$\YGt$\ :& ${\bf 36}$&=&
$({\bf 10,3})\oplus({\bf 6,1})$ \\[.2ex] 
\hline\\[-2ex]
$\YGoo$\ :&  ${\bf 28}$&=&
$({\bf 10,1})\oplus({\bf 6,3})$\\[0.5em]
$\YGtt$\ :& ${\bf 336}$&=&
$({\bf 35,1})
\oplus({\bf 45,3})
\oplus({\bf 20',5})
\oplus({\bf 20',1})
\oplus({\bf 15,3})
\oplus({\bf 1,1})$\\[0.5em]
$\YGoooo$\ :& ${\bf 70}$&=&
$({\bf 20',1})
\oplus({\bf 15,3})
\oplus({\bf 1,5})$\\[1.9ex]
\hline\hline
 & $SU(16)$ & $\supset$ & $SO(10)$ \\ \hline
\Tspan{1ex}{0ex}{$\Onebox75$}\ : & ${\bf 16}$&=&
$({\bf 16})$\\[0.5em]
$\YGt$\ : &
${\bf 136}$&=&
$({\bf \overline{126}})
\oplus({\bf 10})$
\\[.2ex] 
\hline\\[-2ex]
$\YGoo$\ : &  ${\bf 120}$&=&
$({\bf 120})$\\[0.5em]
$\YGtt$\ : &
${\bf 5440}$&=&
$({\bf 4125})
\oplus({\bf \overline{1050}})
\oplus({\bf 54})
\oplus({\bf 210})
\oplus({\bf 1})$\\[0.5em]
$\YGoooo$\ : &
${\bf 1820}$&=&
$({\bf 770})
\oplus({\bf 1050})$\\[1.9ex] \hline
\end{tabular}
\end{center}
\end{table}

\section{Summary and discussions}
\label{Sec:Summary}

We have performed the potential analysis of the $SU(N)$ NJL type 
models for two cases
with a fermion in an $SU(N)$ defining representation $\bfR=\YGo$ and an
$SU(N)$ rank-2 anti-symmetric representation $\bfR=\YIshort\ $,
respectively.

The former case with $\bfR=\YGo$ fermion shows that at the potential
minimum the $SU(N)$ group symmetry is always broken to its special
subgroup $SO(N)$ as far as the symmetry breaking occurs. 
The latter case with $\bfR=\YIshort\ $ also shows that 
the $SU(N)$ symmetry for $N\geq4$ is, if broken, always broken to its 
special subgroup 
$SO(N)$ or $\USp(2[N/2])$ aside from some exceptional cases; 
for $N=4$ the $SU(4)$ symmetry is broken to its special subgroup $SO(4)$
or is not broken although the condensation into $SU(4)$-singlet occurs; 
for $N=16$ the $SU(16)$ is broken to its special subgroup $SO(16)$ or
$SO(10)$ or $\USp(16)$; 
for $N=5$ the $SU(5)$ is broken to its special subgroup $SO(5)$ or to a 
regular subgroup $SU(4)$; 
for $N=3$ the $SU(3)$ is broken to its special subgroup $SO(3)$.
That is, aside from the only breaking $SU(5) \rightarrow SU(4)$ for
$N=5$, all the $SU(N)$ symmetry breakings for $N\geq3$ 
is down to its special subgroups in the case $\bfR=\YIshort\ $.

\begin{figure}[htb]
\begin{center}  
\includegraphics[bb=0 0 34 31, width=50mm]{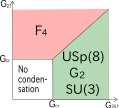}
\end{center}
\caption{$E_6$ phase diagrams are shown in the coupling
 constant $(G_{351^\prime},\,G_{27})$ plane.}
\label{figure:E6-phase-diagram}
\end{figure}

This result clearly shows that symmetry breaking into {\it special}
subgroups is not special at all at least for the dynamical symmetry
breaking in the 4D NJL type model. 
One might, however, suspect that this may be a special situation specific 
to the classical group $G=SU(N)$ model. 
But, actually, this tendency of symmetry breaking to special subgroups 
was found previously for the exceptional group $G=E_6$ model in
Ref.~\cite{Kugo:1994qr}. They analyzed the potential in the 4D $E_6$ NJL
model with fundamental representation $\bfR={\bf27}$ fermion, which have
two coupling constants $G_{\bf27}$ and $G_{\bf351'}$ since
$({\bf27}\times{\bf27})_S = \overline{\bf27}+ \overline{\bf351'}$. The
result of their potential analysis is summarized in the $E_6$ phase
diagram shown in Figure~\ref{figure:E6-phase-diagram}.

This result is very similar to the breaking pattern in our $SU(16)$ case 
with $\YGoo$ fermion shown in Figure~\ref{figure:SU2n-phase-diagram}. 

First of all, all the groups $F_4$, $\USp(8)$, $G_2$ and $SU(3)$
appearing here in Figure~\ref{figure:E6-phase-diagram} are special
subgroups, and the breaking into the regular subgroup $SO(10)=E_5$ does
not occur at all despite that $SO(10)$ is one of the maximal little
groups of scalar $\Phi_{\bf27}$ or $\Phi_{\bf351'}$. 
Moreover, the {\bf27} fermion falls into a single irreducible
representation {\bf 27} under the special subgroups $\USp(8)$, $G_2$ and
$SU(3)$ 
while it splits into two ${\bf26}+{\bf1}$ under $F_4$. This is very much 
parallel to the situation in our $SU(16)$ case that 
the $\YGoo$ fermion {\bf120} falls into a single representation {\bf120} 
also under the $SO(16)$ and $SO(10)$ subgroups, while it splits into 
${\bf119}+{\bf1}$ under $\USp(16)$. In particular, the fact that the
irreducible representation fermion of $G$ also falls into a single
irreducible representation under distinct {\em plural} subgroups $H$
implies in this NJL model the special existence of {\em degenerate
broken vacua}; $\USp(8)$, $G_2$ and $SU(3)$ vacua for the $G=E_6$ case,
and $SO(16)$ and $SO(10)$ vacua for $G=SU(16)$ case.  
For the $E_6$ case, however, numerical study showed the surprising fact 
that the general vacuum does not show any of the symmetries $\USp(8)$, or
$G_2$ or $SU(3)$. The authors of Ref.~\cite{Kugo:1994qr} conjectured the 
existence of the continuous path in the scalar $\Phi_{\bf 351'}$ space 
connecting those three vacua of $\USp(8)$, $G_2$
and $SU(3)$ through which the potential is flat and the $E_6$ symmetry
is totally broken in between those three points. Although this was a 
conjecture for the $E_6$ case, we can show explicitly that it is really 
the case for our $SU(16) \rightarrow SO(16), SO(10)$ breaking case. We
have shown this analytically in Appendix by constructing the
one-parameter vacua which connect the $SO(16)$ and $SO(10)$ vacua and
realize the degenerate lowest potential energy. Explicit computation of
the $SU(16)$ gauge boson mass square matrix was given for the $SO(16)$
and $SO(10)$ vacua which suggests the total breaking of $SU(16)$
symmetry for the general parameter vacua between the $SO(16)$ and
$SO(10)$ vacua.

\section*{Acknowledgment}

This work was supported in part by the MEXT/JSPS KAKENHI Grant Number
JP18K03659 (T.K.), and JP18H05543 (N.Y.)

\appendix

\section{Degeneracy between SO(16) and SO(10) vacua in the SU(16) NJL model}
\label{sec:su16_so16-so10}

As stated in the text, the NJL model with rank-2 anti-symmetric fermion 
$\psi_{\subYGoo}=\psi_I=\psi_{ij}$ for $G=SU(N{=}16)$, 
is broken into the $SO(N{=}16)$-invariant vacuum, when
$G_{\subYGtt}>G_{\subYGoooo}$ as usual for any $N$,  realizing the VEV
\begin{align}
 \VEV{{\Phi_{\subYGtt}\,}_{ij,kl}}^{\rm SO(N{=}16)}=
 \frac{v}2 \delta_{ij}^{kl} 
=v\delta_{[i}^k\delta_{j]}^l \ .
\label{eq:SO16VEV}
\end{align}
For this $N=16$ case, however, $G=SU(16)$ can also be broken to the
SO(10) vacuum possessing the VEV
\begin{equation}
\VEV{{\Phi_{\subYGtt}\,}_{ij,kl}}^{\rm SO(10)}
= \frac{1}{3!3!2^3}(\sigma_{abc}C)_{ij} \frac{v}2\delta_{abc}^{def}(\sigma_{def}C)_{kl},
\label{eq:SO10VEV}
\end{equation}
which also realizes the degenerate mass-square eigenvalue for
$16\cdot15/2=120$ fermions $\psi_{ij}$ determined by the minimum of
$M^2_{\subYGtt}\,x-f(x)$, so realizing the same lowest vacuum energy value
as the above $SO(16)$ vacuum. The $16\times16$ matrix $\sigma_{abc}C$
will be explained shortly below. 

To understand the reason why these two vacua, $SO(16)$ and $SO(10)$, can 
realize the same degenerate {\bf 120} fermion mass-square is interesting
and important, since these two vacua turn out to be continuously
connected with each other via one-parameter family of vacua with
non-vanishing VEV in {\bf 5440} $\Phi_{\subYGtt}$ which all realize the
same degenerate {\bf 120} fermion mass-square but nevertheless violate  
{\em completely} the $SU(16)$ symmetry. 

Similar phenomenon was previously observed in Ref.~\cite{Kugo:1994qr}
which considered the $G=E_6$ NJL model with {\bf 27} fermion: there, the
system has three degenerate broken vacua into $\USp(8)$, $G_2$ and
$SU(3)$, respectively, which all realize the degenerate {\bf 27} fermion
mass-square and hence the lowest vacuum energy for the coupling region
$G_{{\bf351'}}>G_{{\bf 27}}$. The authors of Ref.~\cite{Kugo:1994qr}
performed the numerical search for the potential minimum and actually
found the degenerate mass-square for the {\bf 27} fermion there. But,
they also computed the $E_6$ gauge boson mass eigenvalues on those vacua
to identify the residual unbroken symmetries, and, surprisingly found
that the gauge bosons are all massive and non-degenerate, implying no 
symmetries remain there. They interpreted it that there exist a path in
$\Phi_{\bf 351'}$ space connecting those three vacua of $\USp(8)$, $G_2$
and $SU(3)$ through which the potential is flat and the $E_6$ symmetry
is totally broken in between those three points. This was merely their
interpretation of the numerical results but was not shown
analytically. Here, in this $G=SU(16)$ case, we can show this explicitly
as we now do so.

The $SU(16)$ indices $i, j, \cdots$ taking values
$1, 2, \cdots, N({=}16)$ are identified with the spinor indices of the
special subgroup $SO(10)$. So, it is now necessary to recall some
properties of the $SO(10)$ Clifford algebra, which was explicitly
constructed in the Appendix of Ref.~\cite{Kugo:1994qr}:  Its ten
generators, i.e., ten $32\times32$ gamma matrices $\Gamma_a$ and charge
conjugation matrix ${}^{10}C$ are given in the following form in terms
of the $16\times16$ `Weyl' submatrices $\sigma_a$ and $C$: 
\begin{equation}
\Gamma_a = 
\begin{pmatrix}
0& \sigma_a \\
\sigma_a^\dagger& 0 
\end{pmatrix}\quad (a=1,2,\cdots,10), \quad 
{}^{10}C = 
\begin{pmatrix}
0& C \\
C & 0 
\end{pmatrix}.
\end{equation}
The matrix $C$ is chosen real as 
\begin{equation}
C = 
\begin{pmatrix}
0& -1_4\otimes \epsilon_2 \\
1_4\otimes \epsilon_2 & 0 
\end{pmatrix}
=C^{\rm T}=C^{-1}=C^\dagger, \qquad 
\epsilon_2=
\begin{pmatrix}
0& 1 \\
-1 & 0 
\end{pmatrix},
\end{equation}
and the $16\times16$ $\sigma_a$ matrices satisfy 
\begin{equation}
C\sigma_a^\dagger C^{-1}=\sigma_a^*=
\varepsilon(a)\,\sigma_a \quad \rightarrow\quad 
C\bar\sigma_{abc}=(\sigma_{abc}C)^*=\varepsilon(abc)\,(\sigma_{abc}C)
\label{eq:A5}
\end{equation}
with $\varepsilon(abc):=\varepsilon(a)\varepsilon(b)\varepsilon(c)$,
where the signature factors $\varepsilon(a)$ are $+1$ for five $a$'s and
$-1$ for the other five $a$'s; for the explicit choice of $\sigma_a$ in
Ref.~\cite{Kugo:1994qr}, we have 
\begin{equation}
\varepsilon(a)= 
\begin{cases}
 +1 & \hbox{for}\ a=1,2,3,8,10 \\
 -1 & \hbox{for}\ a=4,5,6,7,9
\end{cases}.
\end{equation}

The anti-symmetric spinor pair index $[ij]$ 
can equivalently be expressed by the rank-3 antisymmetric $SO(10)$ 
vector indices $[abc]$ $(a, b, c, \cdots= 1, 2, \cdots, 10)$ by the 
transformation tensor $(\sigma_{abc}C)_{ij}$ and
$(C\bar\sigma_{abc})^{ij}$, where 
\begin{equation}
\sigma_{abc}=\sigma_{[a}\sigma^\dagger_b\sigma_{c]}, \qquad 
\bar\sigma_{abc}=\sigma^\dagger_{[a}\sigma_b\sigma^\dagger_{c]}.
\end{equation}
This is because the ${}_{10}C_3=120$ matrices $(\sigma_{abc}C)_{ij}$ (or
their complex conjugates $(C\bar\sigma_{abc})^{ij}$) span a complete set
of anti-symmetric $16\times16$ matrices for which exist
$16\cdot15/2=120$ independent ones, and satisfy the completeness
relation: 
\begin{equation}
 \frac1{2^3\cdot 3!}(\sigma_{abc}C)_{ij}(C\bar\sigma_{abc})^{kl} =
  \delta^{kl}_{ij},
\quad 
\frac1{2^3\cdot2!}(\sigma_{abc}C)_{ij}(C\bar\sigma_{def})^{ij} =
\delta_{abc}^{def}.
\end{equation}

Thus our scalar field $\Phi_{\subYGtt\,ij,kl}$ can be equivalently
expressed by 
\begin{align}
\Phi_{\subYGtt\ abc,def}&= \left(\frac1{2!\,2^{3/2}}\right)^2
(C\bar\sigma_{abc})^{ij}(C\bar\sigma_{def})^{kl}\Phi_{\subYGtt\ ij,kl}, 
\label{eq:A9}\\
\Phi_{\subYGtt\ ij,kl}&= \left(\frac1{3!\,2^{3/2}}\right)^2
(\sigma_{abc}C)_{ij}(\sigma_{def}C)_{kl}\Phi_{\subYGtt\ abc,def}.
\label{eq:A10}
\end{align} 
They both possess the same norms:
$\|\Phi_{\subYGtt\ abc,def}\|^2 = 
\|\Phi_{\subYGtt\ ij, kl}\|^2$, where
\begin{align} 
\|\Phi_{\subYGtt\ abc,def}\|^2 &:= 
\frac12\frac1{3!3!}\Phi_{\subYGtt\ abc,def}\cdot \Phi^*_{\subYGtt\ abc,def}, \nn
\|\Phi_{\subYGtt\ ij, kl}\|^2 &:=
\frac12\frac1{2!2!}\Phi_{\subYGtt\ ij, kl}\cdot \Phi_{\subYGtt}^{*\ ij, kl}.
\end{align}

Now, using the relation (\ref{eq:A9}), we can express the $SO(16)$ VEV 
(\ref{eq:SO16VEV}) and $SO(10)$ VEV (\ref{eq:SO10VEV}) in terms of
$\Phi_{\subYGtt\,abc,def}$ of $SO(10)$ rank-3 antisymmetric tensor
basis:
\begin{align}
\VEV{{\Phi_{\subYGtt}\,}_{abc,def}}^{\rm SO(N{=}16)}&=
\frac{v}2 \varepsilon(abc)\,\delta_{abc}^{def} \,,
\label{eq:A12}
\\
\VEV{{\Phi_{\subYGtt}\,}_{abc,def}}^{\rm SO(10)}
&= \frac{v}2\, \delta_{abc}^{def}.
\end{align}  
We can now see that these VEVs are simple {\em diagonal} matrices  
$\propto\delta_{abc}^{def}$ in this $SO(10)$ tensor basis, whose 120
diagonal elements are all $v/2$ for $SO(10)$ vacuum while 60 $v/2$ and
60 $-v/2$ for $SO(16)$ vacuum. The sign factor $\varepsilon(abc)=\pm1$
for the latter in Eq.~(\ref{eq:A12}) came from Eq.~(\ref{eq:A5}) for
rewriting $C\bar\sigma_{abc}$ into $\sigma_{abc}C$ for the $SO(16)$
vacuum. For both vacua, the fermion mass square matrix
$\VEV{{\Phi_{\subYGtt}}}^\dagger\VEV{{\Phi_{\subYGtt}}}$ becomes exactly
the same one $(v/2)^2\delta_{abc}^{def}= (v/2)^2\,1_{120}$ for both
vacua.

Now we can find the one-parameter family of more general vacua
connecting these two vacua: that is, the vacua $|0\rangle^t$
parameterized by $t\in[0,1]$ which realize the scalar field VEV
$\VEV{\Phi_{abc,def}}^t:= \,{}^t\!\bra0 \Phi_{abc,def}\ket0^t$ as
\begin{equation}
\VEV{\Phi_{abc,def}}^t=[\varepsilon(abc)]^t\frac{v}2\,\delta_{abc}^{def}.
\end{equation}
If we introduce a diagonal unitary $120\times120$ matrix $U_t$
\begin{equation}
(U_t)_{abc,\,def} = [\varepsilon(abc)]^{t/2}\delta_{abc}^{def},
\end{equation} 
this VEV can be written as
\begin{equation}
\VEV{\Phi}^t = U_t\, \VEV{\Phi_{\subYGtt}}^{SO(10)}\,U_t^{\rm T}.
\end{equation}
Let us now show that 
\begin{enumerate}
\item 
Although being a unitary matrix, $U_t$ does not belong to the $SU(16)$ 
transformation so that the $G=SU(16)$ symmetry is totally broken on 
the vacua $|0\rangle^t$ for $t\in(0,1)$. 

\item The vacua $|0\rangle^t$ have non-vanishing VEV only in the 
channel $\Phi_{\subYGtt}$\,: 
\begin{equation}
\langle\Phi_{\subYGoooo}\rangle^t=0 \quad \rightarrow\quad \VEV{\Phi}^t=
\VEV{\Phi_{\subYGtt}}^t.
\end{equation}
\end{enumerate}
The first point immediately follows from the fact that the vacuum 
$|0\rangle^t$ is $SO(10)$ vacuum at $t=0$ and $SO(16)$ vacuum at $t=1$. 
That is, the isometry group changes as $t$ changes, while the isometry
group cannot change if $U_t$ is an $SU(16)$ transformation. 

The second point is proved as follows. Since the general $120\times120$ 
symmetric matrix $\Phi$ is decomposed into two irreducible components, 
$\Phi_{\subYGoooo}$ and $\Phi_{\subYGtt}$, it is sufficient to show that 
$\Phi_{\subYGoooo}$ component is vanishing on the vacua $\ket0^t$, which 
is given by
\begin{align}
\langle\Phi_{\subYGoooo\,ij,kl} \rangle^t
&\propto\delta^{i'j'k'l'}_{ijkl}\sum_{a,b,c} (\sigma_{abc}C)_{i'j'}\,[\varepsilon(abc)]^t (\sigma_{abc}C)_{k'l'} =  A_{ijkl} + (-1)^t\, B_{ijkl},
\nn
A_{ijkl}&= \delta^{i'j'k'l'}_{ijkl}\sum_{(a,b,c) \text{ with } \varepsilon(abc)=+1} (\sigma_{abc}C)_{i'j'}\,(\sigma_{abc}C)_{k'l'}, \nn
B_{ijkl}&= \delta^{i'j'k'l'}_{ijkl}\sum_{(a,b,c) \text{ with } \varepsilon(abc)=-1} (\sigma_{abc}C)_{i'j'}\,(\sigma_{abc}C)_{k'l'},
\end{align}
where $A_{ijkl}$ and $B_{ijkl}$ are the sum over the 60 sets of 
$(a, b, c)$ with $\varepsilon(abc)=\pm1$, 
respectively. We already know that $\VEV{\Phi}^t$ belongs to
$\Phi_{\subYGtt}$ at the end points $t=0$ and $t=1$, so we have
\begin{equation}
\begin{cases}
t=0 &\rightarrow\ A_{ijkl}+ B_{ijkl} =0  \\  
t=1 &\rightarrow\ A_{ijkl}- B_{ijkl} =0 
\end{cases}
\quad \rightarrow\quad  A_{ijkl}=B_{ijkl} =0.
\end{equation}
Thus, $\langle\Phi_{\subYGoooo} \rangle^t$ vanishes for any $t$,
proving the second point. 

This property $\langle\Phi_{\subYGoooo} \rangle^t=0$ guarantees that 
all the vacua $\ket0^t$ realize the lowest energy states degenerate with
the $SO(10)$ and $SO(16)$ vacua at the endpoints $t=0$ and $t=1$; this
is because the potential is commonly calculated by 
$M_{\subYGtt}^2\tr(\Phi^\dagger\Phi)- \tr f(\Phi^\dagger\Phi)$ 
since $\Phi=\Phi_{\subYGtt}$ for these vacua. 

\subsection{Mass square matrix of $SU(16)$ gauge boson}

In order to see which symmetry actually remains on a vacuum with given
VEV, one way is to see the mass spectrum of the gauge boson for (gauged)
$G$ symmetry. It is also necessary to calculate the gauge boson masses
in order to see how the degeneracy of the vacuum energy due to the
fermion loop is lifted by the gauge boson loop contribution. 

It is actually difficult to analytically calculate the gauge boson mass 
square matrix for the general vacua $\ket0^t$ given above, since all 
the $G=SU(16)$ symmetry is expected lost there. So we calculate it 
only at the two end points, $SO(16)$ and $SO(10)$ vacua and guess the 
spectrum by interpolation. 

The scalar kinetic term $(D_\mu\Phi)^\dagger(D^\mu\Phi)$ gives 
the gauge boson mass term $(1/2)M_{AB}^2A_\mu^AA^{B\,\mu}$ by substituting 
the VEV for the scalar field $\Phi$. Since the derivative term
$\partial_\mu\Phi$ does not contribute for the constant VEV, this
implies that we can find the mass square matrix $M_{AB}^2$ by simply
calculating the square of the gauge transformation $\delta(\theta)$: 
\begin{equation}
\|\delta(\theta)\Phi\|^2 = \frac12 \theta^A\theta^B M_{AB}^2.
\end{equation}
The $G=SU(N=16)$ transformation for this case is given by 
\begin{equation}
\delta(\theta)\Phi_{ij,kl}
= 2\Theta_{[i}{}^{i'}\delta_{j]}{}^{j'}\Phi_{i'j',kl} + 
\Bigl( (i,j) \leftrightarrow (k,l) \Bigr), \quad 
\label{eq:SUtrf}
\end{equation}
where 
\begin{align}
\Theta_i{}^j &= \sum_{A=1}^{N^2-1} g\theta^A(T_A)_i{}^j \nn
 &= \frac1{2!}g\phi^{ab}(T_{ab})_i{}^j
 + \frac1{4!}g\theta^{abcd}(T_{abcd})_i{}^j.
\label{eq:SUgens}
\end{align}
Here $g$ is the gauge coupling and the second line is 
the particular choice of the $SU(16)$ generators 
respecting the $SO(10)$ subgroup: $SU(16)$ adjoint
${\bf255}= {\bf45}+{\bf210}$ of $SO(10)$. We adopt the convention 
$\tr(T_A T_B)= (1/2)\delta_{AB}$ for Hermitian generators
$T_A^\dagger=T_A$, then, 
\begin{align}
T_{ab}&= \frac{i}{\sqrt{2N}}\sigma_{ab}, \qquad 
 \Bigl(\sigma_{ab}=\sigma_{[a}\sigma^\dagger_{b]},
 \quad \frac1{2^4}\tr(\sigma_{ab}\sigma_{cd})=\delta_{ab}^{cd}\Bigr), \\
T_{abcd}&= \frac{1}{\sqrt{2N}}\sigma_{abcd}, \qquad 
\Bigl(\sigma_{abcd}=\sigma_{[a}\sigma^\dagger_{b}\sigma_{c}\sigma^\dagger_{d]}, \quad 
\frac1{2^4}\tr(\sigma_{abcd}\sigma_{efgh})=\delta_{abcd}^{efgh} \Bigr).
\end{align}

The $G=SU(16)$ transformation on the $SO(16)$ vacuum is most easily 
computed by using the VEV (\ref{eq:SO16VEV}), 
$\VEV{\Phi_{ij,kl}}^{SO(16)}=(v/2)\delta_{ij}^{kl}$:
\begin{align}
\VEV{\delta(\theta)\Phi_{ij,kl}}^{SO(16)}  
&= 2\Theta_{[i}{}^{i'}\delta_{j]}{}^{j'}\VEV{\Phi_{i'j',kl}}^{SO(16)} + 
\bigl( (i,j) \leftrightarrow (k,l) \bigr) \nn
&=2\cdot\frac{v}2 \left(
2\Theta_{[i}{}^{[k}\delta_{j]}{}^{l]} + 
\bigl( (i,j) \leftrightarrow (k,l) \bigr)
\right)
=2v 
(\Theta+\Theta^{\rm T})_{[i}{}^{[k}\delta_{j]}{}^{l]}.
\end{align}
The norm square is computed as
\begin{align}
\|\VEV{\delta(\theta)\Phi_{ij,kl}}^{SO(16)}\|^2
&=\frac1{2\,2!2!}(2v)^2
(\Theta+\Theta^{\rm T})_{i}{}^{k}\delta_{j}{}^{l}
(\Theta+\Theta^{\rm T})_{[k}{}^{[i}\delta_{l]}{}^{j]} \nn
&=\frac1{2\,2!2!}(2v)^2\frac14 \left[
(N-2) \tr(\Theta+\Theta^{\rm T})^2 + \
\bigl( \tr(\Theta+\Theta^{\rm T})\bigr)^2\right] 
\end{align}
with $N=16$. Note that $\Theta+\Theta^{\rm T}=
2g\theta_{\text{S}}^AT_{\text{S}A}$ is given by the sum only over the
{\em symmetric} matrices $T_{\text{S}A}$, which stand for the broken
generators for $G=SU(16)\rightarrow SO(16)$ and recall that the
generators of unbroken $SO(16)$ consist of all the {\em antisymmetric}
$N\times N$ matrices whose dimension is $N(N-1)/2=120$. Using also
$\tr\Theta=\tr\Theta^{\rm T}=0$ for $SU(N)$ case, and
$\tr(T_{\text{S}A}T_{\text{S}B})=(1/2)\delta_{AB}$, we find 
\begin{align}
\|\VEV{\delta(\theta)\Phi_{ij,kl}}^{SO(16)}\|^2
&=\frac14 g^2v^2 (N-2) \sum_{A,B\in SU(16)/SO(16)}\theta_{\text{S}}^A\theta_{\text{S}}^B\delta_{AB}.
\end{align}
Namely, the gauge bosons for the $SO(16)$ {\bf 135} broken generators 
$\in SU(16)/SO(16)$ get a 
common mass square 
\begin{equation}
M^2_{SU(16)/SO(16)}=\frac{7}{2}g^2v^2,
\end{equation}
while gauge bosons for the unbroken 45 $SO(16)$ generators of course 
remain massless.

Next compute the gauge boson masses for the $SO(10)$ vacuum case, for
which the VEV is simpler in the $SO(10)$ vector index basis: 
\begin{equation}
\VEV{\Phi_{\subYGtt\,abc,def}}^{SO(10)}= \frac{v}2\delta_{abc}^{def}.
\end{equation}
So the computation is simpler if we first convert the $G=SU(16)$
transformation law (\ref{eq:SUtrf}) in $SU(16)$ spinor index $i,j$ basis
into that in $SO(10)$ vector index $a,b,c$ basis by using the conversion
formula (\ref{eq:A9}) and (\ref{eq:A10}):
\begin{equation}
\delta(\theta)\Phi_{abc, def}
 = -2\frac1{2!\,3!\,2^3}
 \tr\bigl(C\bar\sigma_{abc}\Theta\sigma_{a'b'c'}C\bigr)\,
\Phi_{a'b'c',def} + 
\Bigl( (a,b,c) \leftrightarrow (d,e,f) \Bigr).
\label{eq:SUtrfabc}
\end{equation}
Using the fusion rule for the gamma matrices
\begin{equation}
\sigma_{abc}\bar\sigma^{def}= \sigma_{abc}{}^{def}
+9\delta_{[a}^{[d}\sigma_{bc]}{}^{ef]}
-9\delta_{[ab}^{[de}\sigma_{c]}{}^{f]}- \delta_{abc}^{def}
\end{equation}
and the $SO(10)$ decomposition (\ref{eq:SUgens}) of the $SU(16)$
transformation parameter,
$\Theta=\phi^{ab}\sigma_{ab}/2! + \theta^{abcd}\sigma_{abcd}/4!$, we find
\begin{align}
&\frac1{2^4}\tr\bigl(C\bar\sigma_{abc}\Theta\sigma_{a'b'c'}C\bigr)
=\frac1{2^4}\tr\bigl(\Theta\sigma_{a'b'c'}\bar\sigma_{abc}\bigr) \nn
& \hspace{2em}{}=  \frac{g}{\sqrt{2N}}\left( 
-\frac1{4!}\epsilon_{abc\,a'b'c'\,\alpha\beta\gamma\delta}\theta^{\alpha\beta\gamma\delta} 
+ 9 \delta_{[a'}^{[a}\theta_{b'c']}{}^{bc]} 
+ 9i \delta_{[a'b'}^{[ab}\phi_{c']}{}^{c]} 
\right).
\end{align}
Substituting this into (\ref{eq:SUtrfabc}) and taking the VEV on the 
$SO(10)$ vacuum, we find
\begin{align}
\VEV{\delta(\theta)\Phi_{abc, def}}^{SO(10)}& \nn
= -\frac{2g}{3!\sqrt{2N}} \biggl( 
\Bigl[
&-\frac1{4!}\epsilon_{abc\,a'b'c'\,\alpha\beta\gamma\delta}\theta^{\alpha\beta\gamma\delta}
\cdot\frac{v}2 \delta^{a'b'c'}_{def}  
+ 9 \delta_{[a'}^{[a}\theta_{b'c']}{}^{bc]} 
\cdot\frac{v}2 \delta^{a'b'c'}_{def} \nn
&{}+ 9i \delta_{[a'b'}^{[ab}\phi_{c']}{}^{c]} 
\cdot\frac{v}2 \delta^{a'b'c'}_{def} 
\Bigr] +
\Bigl[ (a,b,c) \leftrightarrow (d,e,f) \Bigr]
\biggr) \nn
= -\frac{2g}{\sqrt{2N}}\frac{v}2 \biggl( 
\Bigl[
&-\frac1{4!}\epsilon_{abc\,def\,\alpha\beta\gamma\delta}\theta^{\alpha\beta\gamma\delta}
+ 9 \delta_{[d}^{[a}\theta_{ef]}{}^{bc]} \nn
&{}+ 9i \delta_{[de}^{[ab}\phi_{f]}{}^{c]} 
\Bigr] +
\Bigl[ (a,b,c) \leftrightarrow (d,e,f) \Bigr]
\biggr).
\label{eq:A33}
\end{align}
In the last expression, the factors
$\epsilon_{abc\,def\,\alpha\beta\gamma\delta}$ and 
$\delta_{[de}^{[ab}\phi_{f]}{}^{c]}$ are seen to be anti-symmetric 
under the exchange $(a,b,c) \leftrightarrow (d,e,f)$, so the 
first and the third terms in the first square bracket 
are canceled by the $(a,b,c) \leftrightarrow (d,e,f)$ 
exchanged terms while the second term is doubled. Thus, finally, we obtain
\begin{equation}
\VEV{\delta(\theta)\Phi_{abc, def}}^{SO(10)}
= -\frac{gv}{\sqrt{2N}} 9 \delta_{[d}^{[a}\theta_{ef]}{}^{bc]}\times2.
\end{equation}
The norm square is calculated as
\begin{align}
\|\VEV{\delta(\theta)\Phi_{abc, def}}^{SO(10)}\|^2 
&= \frac1{2\,3!\,3!}\frac{g^2v^2}{2N}2^2 9 \,
\delta_{d}^{a}\theta_{ef}{}^{bc}\cdot 9\,\delta_{[a}^{[d}\theta_{bc]}{}^{ef]}.
\end{align}
Expanding 
\begin{align}
\delta_{d}^{a}\theta_{ef}{}^{bc}&\cdot 9\,\delta_{[a}^{[d}\theta_{bc]}{}^{ef]}
=\theta_{ef}{}^{bc}\cdot 9\,\delta_{[a}^{[a}\theta_{bc]}{}^{ef]} \nn
&=\theta_{ef}{}^{bc}\Bigl(
3 \delta_{a}^{[a}\theta_{bc}{}^{ef]}
+3 \delta_{b}^{[a}\theta_{ca}{}^{ef]}
+3 \delta_{c}^{[a}\theta_{ab}{}^{ef]} 
\Bigr) \nn
&=\theta_{ef}{}^{bc}\Bigl(
(10{-}1{-}1)+({-}1{+}0{+}0)+({-}1{+}0{+}0)\Bigr)\theta_{bc}{}^{ef}
= 6\,\theta^{abcd}\theta_{abcd},
\end{align}
we have
\begin{align}
\|\VEV{\delta(\theta)\Phi_{abc, def}}^{SO(10)}\|^2 
&= \frac1{2\,3!\,3!}\frac{g^2v^2}{2N}2^2 9\times 
6\,\theta^{abcd}\theta^{abcd} 
=\frac94g^2v^2 \bigl(
\frac1{4!}\theta^{abcd}\theta^{abcd}\bigr).
\end{align}
This tells us that the $SO(10)$ {\bf210} gauge bosons corresponding to
the broken generators $T_{abcd}$ $\in SU(16)/SO(10)$ get a common mass
square 
\begin{equation}
M^2_{SU(16)/SO(10)}= \frac{9}{2} g^2v^2,
\end{equation}
while the other $SO(10)$ {\bf45} gauge bosons for the unbroken generators 
$T_{ab}$ remain massless. 

Finally, two comments are in order: First, for the interpolating vacua 
$\ket0^t$, the gauge transformation (\ref{eq:A33}) is replaced by
\begin{align}
\VEV{\delta(\theta)\Phi_{abc, def}}^t
= -\frac{2g}{\sqrt{2N}}\frac{v}2 \biggl( 
\Bigl[
 &-\frac1{4!}\epsilon_{abc\,def\,\alpha\beta\gamma\delta}
 \theta^{\alpha\beta\gamma\delta}[\varepsilon(def)]^t
+ 9 \delta_{[d}^{[a}\theta_{ef]}{}^{bc]}[\varepsilon(def)]^t \nn
&{}+ 9i \delta_{[de}^{[ab}\phi_{f]}{}^{c]} [\varepsilon(def)]^t
\Bigr] +
\Bigl[ (a,b,c) \leftrightarrow (d,e,f) \Bigr]
\biggr).
\label{eq:A39}
\end{align}
Then, the cancellation between the $(a,b,c) \leftrightarrow (d,e,f)$ 
exchanged terms no longer occur and it seems that all the generators are
broken. The $[\varepsilon(def)]^t$ factors do not cancel in the
computation of norm square, and the analytical calculation becomes very
complicated. 

Second comment is on the gauge boson 1-loop contribution to the vacuum
energy as a perturbation. Since the boson 1-loop contribution of mass
$m$ is expected to be $+f(m^2)$, the $SO(16)$ and $SO(10)$ vacua have
the following additional contribution to the degenerate vacuum energy:
\begin{align}
SO(16) \text{vacuum}:\qquad  &135 f(\frac{14}{2} g^2v^2), \nn
SO(10) \text{vacuum}:\qquad  &210 f(\frac{9}{2} g^2v^2).
\end{align}      
Since the total sum of gauge boson mass squares is the same between the
two vacua, $135\times14= 210\times9=1890$, the upward convexity of the
function $f(x)$ leads to the inequality,
$210 f(\frac{9}{2}g^2v^2)> 135 f(\frac{14}{2} g^2v^2)$. This implies that the
gauge boson 1-loop contribution lifts the degeneracy between the two
vacua $SO(16)$ and $SO(10)$, and $SO(16)$ vacuum will be realized as the 
lowest energy vacuum.

\bibliographystyle{utphys} 
\bibliography{../../arxiv/reference}

\end{document}